\documentclass{article}

\usepackage[preprint]{neurips_2024}

\usepackage[utf8]{inputenc} 
\usepackage[T1]{fontenc}    
\usepackage{hyperref}       
\usepackage{url}            
\usepackage{booktabs}       
\usepackage{amsfonts}       
\usepackage{nicefrac}       
\usepackage{microtype}      
\usepackage{xcolor}         
\usepackage{amsmath}
\usepackage{amssymb}
\usepackage{mathtools}
\usepackage{amsthm}
\usepackage{bm}
\usepackage{multirow}
\usepackage{wrapfig}
\theoremstyle{plain}
\newtheorem{theorem}{Theorem}[section]

\newtheorem{lemma}[theorem]{Lemma}
\newtheorem{corollary}[theorem]{Corollary}
\newtheorem{definition}[theorem]{Definition}

\theoremstyle{remark}

\usepackage{bbding}

\newcommand{\x}{\bm{x}}

\renewcommand{\k}{\bm{k}}

\newcommand{\sk}{\textup{\textsf{sk}}}

\renewcommand{\k}{\bm{k}}

\newcommand{\cF}{\mathcal{F}}

\newcommand{\cV}{\mathcal{V}}

\newcommand{\cP}{\mathcal{P}}

\newcommand{\bbE}{\mathbb{E}}

\newcommand{\bbR}{\mathbb{R}}

\usepackage{algorithm}
\usepackage[noend]{algpseudocode}
\title{Distortion-free Watermarks are not Truly Distortion-free under Watermark Key Collisions}
\author{
Yihan Wu$^{1}$\thanks{Equal contribution}\quad Ruibo Chen$^{1*}$\quad Zhengmian Hu$^{1}$\quad Yanshuo Chen $^{1}$ \\
\textbf{Junfeng Guo} $^{1}$\quad \textbf{Hongyang Zhang}$^{2}$\quad \textbf{Heng Huang}$^{1}$ \\
$^{1}$ University of Maryland, 
$^{2}$ University of Waterloo
}

\begin{document}

\maketitle

\begin{abstract}
Language model (LM) watermarking techniques inject a statistical signal into LM-generated content by substituting the random sampling process with pseudo-random sampling, using watermark keys as the random seed. Among these statistical watermarking approaches, distortion-free watermarks are particularly crucial because they embed watermarks into LM-generated content without compromising generation quality. 
However, one notable limitation of pseudo-random sampling compared to true-random sampling is that, under the same watermark keys (i.e., \textit{key collision}), the results of pseudo-random sampling exhibit correlations. This limitation could potentially undermine the distortion-free property.
Our studies reveal that key collisions are inevitable due to the limited availability of watermark keys, and existing distortion-free watermarks exhibit a significant distribution bias toward the original LM distribution in the presence of key collisions. Moreover, achieving a perfect distortion-free watermark is impossible as no statistical signal can be embedded under key collisions. To reduce the distribution bias caused by key collisions, we introduce a new family of distortion-free watermarks--beta-watermark. 
Experimental results support that the  beta-watermark can effectively reduce the distribution bias under key collisions. Code is available at\footnote{https://github.com/RayRuiboChen/RethinkingWatermark}.
\end{abstract}

\vspace{-0.2cm}
\section{Introduction}
\vspace{-0.2cm}
In an era where artificial intelligence surpasses human capabilities in generating text, the authenticity and origin of such AI-generated content have become paramount concerns. Language model watermarking \citep{Aaronson2022,kirchenbauer2023watermark,christ2023undetectable,kuditipudi2023robust,hu2023unbiased} provides a promising solution for distinguishing between human and machine-generated text.  This technique secretly embeds a statistical signal into the generated text using a pseudo-random generator seeded with watermark keys. The embedded signal is then detected through a statistical hypothesis test, ensuring the traceability and verification of the text's origin.

Distortion-free watermarks \citep{Aaronson2022,christ2023undetectable,kuditipudi2023robust,hu2023unbiased} represent one of the most compelling techniques in language model watermarking. These watermarks are particularly valuable because they provably preserve the output distribution of the original language model. Specifically, the expected watermarked distribution with respect to the watermark keys remains identical to the original language model distribution, thus offering significant practical application potential.

However, the pseudo-random nature of the watermark generator may lead to correlations between generated content when the watermark keys are identical (i.e., key collision). In extreme cases, such as when the prompt remains the same, key collisions can result in identical generated content, significantly limiting its application scenarios. For instance, when using GPT-4 to generate content, if the initial output is unsatisfactory, a request to regenerate would typically yield a different result. However, under a distortion-free watermarking scheme, the output may remain unchanged due to the consistent application of the same watermark key. This limitation highlights a critical challenge in the practical deployment of such watermarking techniques.



In our research, we comprehensively analyze the existing distortion-free watermarks and demonstrate, through both theoretical and empirical evidence, that \emph{no distortion-free watermark can fully preserve the original LM distribution under key collisions}. Specifically, we categorize the level of distortion-free capability into three types:
a) Step-wise distortion-free---the watermark preserves the LM distribution at a single token generation step;
b) Weakly distortion-free---the watermark preserves the LM distribution for a one-time sentence generation;
c) Strongly distortion-free---the watermark preserves the LM distribution across multiple sentence generations.
Our findings indicate that all existing distortion-free watermarks are weakly distortion-free but not strongly distortion-free due to key collisions. In particular, we theoretically prove that there does not exist any strong distortion-free watermark under key collisions. We also show that key collisions are inevitable given the limited number of watermark keys available in current schemes.

To mitigate the distribution bias caused by key collisions, we introduce the beta-watermark and develop a novel model-agnostic detector that can identify watermarks without requiring access to prompts or language models. Additionally, we design empirical metrics to measure the distribution bias resulting from key collisions. Through rigorous testing on widely-studied language models, including BART-large model~\citep{liu2020multilingual} and LLaMA-2~\citep{touvron2023llama2}, our beta-watermark has demonstrated effectiveness in significantly reducing the distribution bias induced by key collisions.


Our main contributions are summarized as follows:
\begin{itemize}
\item We identify three levels of distortion-free capabilities in watermarks—Step-wise, Weakly, and Strongly Distortion-free—revealing that existing watermarks are not strongly distortion-free and cannot preserve the original language model distribution under multiple generations due to the inevitability of key collisions. 

\item Under watermark key collisions, we theoretically demonstrate a trade-off between watermark strength and its distribution bias to the original LM distribution—a smaller distribution bias results in weaker watermark strength. Especially, we prove that the distribution bias of a strongly distortion-free watermark is always zero, corresponding to zero watermark strength. This suggests that strongly distortion-free watermarks do not exist under key collisions.

\item We introduce beta-watermark, a new family of weakly distortion-free watermarks that can provably reduce the distribution bias caused by key collisions. Besides, we design a novel model-agnostic detector that identifies watermarks without needing access to prompts or specific language models.
Through experiments on popular language models like BART-large and LLaMA-2, we demonstrate our theoretical findings that existing watermarks are not strongly distortion-free and beta-watermark can effectively reduce the distribution bias.
\end{itemize}

\vspace{-0.2cm}
\section{Related Work}
\vspace{-0.2cm}
\textbf{Statistical watermarks.} \cite{kirchenbauer2023watermark} enhanced the statistical watermark framework originally introduced by \cite{Aaronson2022}, demonstrating the effectiveness of statistical watermarking through extensive experiments on large language models. 
They splited the LM tokens into red and green list, then promoted the use of green tokens by adding a fixed parameter $\delta$ to their logits. \cite{zhao2023provable} proposed the unigram watermark, which enhances the robustness of the statistical watermark by using one-gram hashing to produce watermark keys. \cite{liu2023semantic} also improved the robustness of statistical watermarking by leveraging the semantics of generated content as watermark keys. Additionally, \cite{liu2023unforgeable} proposed an unforgeable watermark scheme that employs neural networks to modify token distributions instead of using traditional watermark keys. However, these approaches may lead to significant changes in the distribution of generated text, potentially compromising content quality.

\textbf{Distortion-free watermarks.} To preserve the original output distribution in watermarked content, researchers have explored alternative strategies to modify the token distribution. \cite{Aaronson2022} introduced the first distortion-free watermarking strategy, which utilized Gumbel-reparametrization to alter token distribution and the prefix n-gram content as the watermark keys. \cite{christ2023undetectable} and \cite{kuditipudi2023robust} adopted the inverse-sampling and Gumbel-reparametrization to modify the watermarked token distributions, where the watermark keys are based on the token position or a fixed key list respectively. Notice \cite{christ2023undetectable}'s method encounters resilience challenges under modifications and lacks empirical evidence regarding its detectability. Meanwhile, \cite{kuditipudi2023robust}'s detection process involves hundreds of resampling steps from the secret key distribution, proving inefficient for processing lengthy texts. \cite{hu2023unbiased} employed inverse-sampling and permute-reweight methods for watermarking. But their detector is not model-agnostic, which requires access to the language model API and prompts, which compromises its operational efficiency. A detailed related work section is in Appendix~\ref{sec:add related work}.

\vspace{-0.2cm}
\section{Preliminary} 
\vspace{-0.2cm}
\textbf{Notations.} 
Denote by $V:=\{t_1,...,t_N\}$ the vocabulary (or token) set of a language model, and by $N = |V|$ its size. Let $\cV$ represent the set of all conceivable string sequences, including those of zero length. A language model generates a token sequence based on a predetermined prompt. For a single step in this process, the probability of generating the next token $x_{n+1} \in V$, given the current context from $x_1$ to $x_n$, is represented as $P_M(x_{n+1} \mid x_1, x_2, \dots, x_n)$. For brevity, we adopt the condensed notation: 
$P_{M}(\bm{x}_{n+1:n+m} \mid \bm{x}_{1:n})$, where $\bm{x}_{n+1:n+m} = (x_{n+1}, \dots, x_{n+m})$. Note that the prompt is deliberately omitted in this representation. Inherent to its design, the language model operates in an autoregressive mode. This implies that the combined probability of generating several tokens, specifically from $x_{n+1}$ to $x_{n+m}$, takes the form 
$P_{M}(\bm{x}_{n+1:n+m} \mid \bm{x}_{1:n}) = \prod_{i=1}^m P_{M}(x_{n+i} \mid \bm{x}_{1:n+i-1}).$

\textbf{Watermarking problem definition.} A language model (LM) service provider aims to watermark the generated content such that all other users can verify if the content is generated by the LM without needing access to the LM or the original prompt. A watermark framework primarily consists of two components: a \textit{watermark generator} and a \textit{watermark detector}. The watermark generator embeds a watermark into the text through a \textit{Pseudo-random Distribution Adjustment rule} (PDA-rule), which is seeded by watermark keys. The watermark detector, on the other hand, detects the presence of the watermark within the content using a statistical hypothesis test.

\begin{definition}[PDA-rule] Let $\cP$ represent the space of token distributions and let $K$ denote the space of watermark keys. A Pseudo-random Distribution Adjustment rule (\textbf{PDA-rule}), defined as $F: \cP \times K \to \cP$, adjusts the token distribution based on a given watermark key.
\end{definition}
\textbf{Watermark generator.} During the watermark generation process, the service provider modifies the original language model distribution $P_M$ using a \textit{watermark key} $k \in K$ and a PDA-rule. Here, the watermark key acts as a random seed to modify the distribution, after which the next token is sampled from this modified distribution. A watermark key usually consists of a \textit{secret key} $\sk$ and a context key (e.g., n-gram \citep{Aaronson2022} or token position \citep{christ2023undetectable}). Let $\cF := \{F: \cP \times K \to \cP\}$ denote the set of PDA-rules. Specifically, let $P_W$ denote the distribution of the LM after watermarking, and $k$ the watermark key, $P_W(t \mid \bm{x}_{1:n-1}) := F(P_M(\cdot \mid \bm{x}_{1:n-1}), k)(t),\forall t\in V$, where $P_M(\cdot \mid \bm{x}_{1:n-1})$  is the LM token distribution for sampling the $n$-th token. When sampling the next token $x_n$, the language model samples from $P_W(\cdot \mid \bm{x}_{1:n-1})$ instead of $P_M(\cdot \mid \bm{x}_{1:n-1})$. This mechanism allows the service provider to inject a statistical signal into the generated content.

The PDA-rule is the core of the watermark generator. A PDA-rule is considered distortion-free if and only if it preserves the token distribution during watermark generation. To the best of our knowledge, there are three types of distortion-free PDA-rules: inverse-sampling \citep{christ2023undetectable,kuditipudi2023robust,hu2023unbiased}, Gumbel-reparametrization \citep{Aaronson2022,kuditipudi2023robust}, and permute-reweight \citep{hu2023unbiased}. A detailed introduction to these methods can be found in Section~\ref{sec:exist pda rule}. The formal definition of a distortion-free PDA-rule is presented below.
\begin{definition}[Distortion-free PDA-rule] A PDA-rule $F$, is a distortion-free PDA-rule, if and only if for an arbitrary LM $P_M$,  $\forall \x_{1:n} \in \cV$, and  $\forall i \leq n$, it holds that $P_M(x_i|\x_{1:i-1}) = \bbE_{k_i}[F(P_M(\cdot|\x_{1:i-1}),k_i)(x_i)]$.
\end{definition}

\textbf{Watermark Detector.} During the process of watermark detection, the user will have access only to the watermark key and the PDA-rule $F$. The detector employs a hypothesis testing approach to identify the presence of the watermark signal. The hypothesis test is defined as: $H_0:$ \textit{The content is generated without the presence of watermarks}, and 
$H_1:$ \textit{The content is generated with the presence of watermarks}. For the purposes of the statistical test, a score function $s(x,k,F): V \times K \times \mathcal{F} \to \mathbb{R}$ is employed. Under $H_0$, the score function is a random variable $S_{H_0}$ where $\Pr(S_{H_0} = s(t,k,F)|k,F)=\sum_{s(t',k,F) =s(t,k,F) }P_M(t'),\forall t\in V$, while under $H_1$, the random variable $S_{H_1}$ becomes $\Pr(S_{H_1} = s(t,k,F)| k, F)=\sum_{s(t',k,F) =s(t,k,F) }P_W(t')$. Thus, we can use the discrepancy between $S_{H_0}$ and $S_{H_1}$ to detect the watermark content. Given an observation (text sequence) $\x_{1:n}$, we define the test statistic $S(\x_{1:n}) = \sum_{i=1}^n s(x_i,k,F)$ as the measure for the test. The decision to reject the null hypothesis is based on the difference between $S(\x_{1:n})$ and the expected value $\mathbb{E}_{H_0}[S(\x_{1:n})]$.




\textbf{Watermark Key.} For each generating step, we will use a watermark key to seed the PDA-rule. There are generally three key sampling methods:
\begin{itemize}
    \item \textbf{(n-gram hashing)} \cite{Aaronson2022} and \cite{hu2023unbiased} use a fixed \textbf{secret key} $\sk_0$ and the prefix n-gram $s$ (e.g., $s = \mathbf{x}_{l-n:l-1}$ for generating $x_l$) to form the watermark keys, i.e., $K = \{(\sk_0,s) \mid s \in \mathcal{V}_n\}$, where $\mathcal{V}_n$ represents the set of all n-grams with token set $V$. A history list is kept during one generation to ensure the watermark keys are unique. If the length of previously generated tokens is less than $n$, all preceding tokens are used as $s$.
    \item \textbf{(position hashing)} \cite{christ2023undetectable} uses a fixed \textbf{secret key} $\sk_0$ and the token position are used as watermark keys, i.e., $K = \{(\sk_0,i) \mid i \in \mathbb{N}\}$.
    \item \textbf{(fixed key set)} \cite{kuditipudi2023robust} uses a fixed \textbf{secret key} $\sk_0$ generates a set of watermark keys, $K = \{k_1, \ldots, k_{n_0}\}$. During token generation at step $i$, a random integer $r$ is sampled, and $k_{(i+r) \mod n_0}$ is used as the seed for the PDA-rule. If the token length exceeds $n_0$, we will sample from the original LM distribution instead.
\end{itemize}
\begin{definition}[Key collision] Key collision refers to scenarios where the same watermark keys are used to seed the PDA-rule.
\end{definition}
All three watermark key sampling methods mentioned previously have a limited number of keys given the fixed secret key $\sk_0$. The maximum key volume is $|V|^n$ for n-gram hashing, $l_0$ for position hashing, and $n_0$ for the fixed key set. Here, $l_0$ represents the maximum token length for the language model, typically ranging from $10^4$ to $10^6$. Therefore, if we only have one secret key, key collisions will occur when the number of queries and the generated tokens exceeds the key volume.


\vspace{-0.2cm}
\section{Curse of Key Collision on Distortion-Free Watermarks}
\vspace{-0.2cm}
We start with showing key collision is inevitable. In the previous section, we show that given a fixed secret key $\sk_0$, the watermark key space is finite. Consequently, key collisions will occur with a sufficient number of queries to the language model. One might naturally question whether using an infinite number of secret keys (e.g., a unique key for each generation) could expand the watermark key space to infinity, thereby reducing the likelihood of collisions. However, this approach is impractical because it would substantially reduce detection efficiency. When analyzing a watermarked sequence, the detection algorithm would need to be applied to all possible secret keys, even though only one key corresponds to the watermark. Thus, the watermark information becomes obscured by the numerous other keys. All missing proofs can be found in Appendix~\ref{sec:missing proof}.
\begin{theorem}[Detection efficiency with multiple secret keys]\label{thm:multikey detection} Denote by $S(\cdot|\sk)$ the test statistic. Under the null hypothesis $H_0$, given a random text $\x_{1:n}$, we have $\Pr(S(\x_{1:n}|\sk_0)-\bbE_{H_0}[S]\geq t|H_0)= p_0(t),$ i.e., $p_0(t)$ is the false positive rate of threshold $t$ under single secret key detection. Given $M$ different secret keys, if we use the maximum of the score as the test statistic, we have
    $$\Pr\left(\max_{i\in[M]}(S(\x_{1:n}|\sk_i)-\bbE_{H_0}[S])\geq t|H_0\right)= 1-(1-p_0(t))^M,\quad \forall t\in\bbR.$$
\end{theorem}
\begin{corollary}Under the existing watermark key sampling schemes, key collision is inevitable.
\end{corollary}
    Theorem~\ref{thm:multikey detection} states that, given the same threshold $t$, the false positive rate increases with the number of secret keys. Especially, when $M\to\infty$, the false positive rate will tend to 1, which indicates every sentence will be detected as watermarked. Thus, the number of secret keys should be finite, and key collision is inevitable.

We then provide the definition of the three levels of distortion-free capabilities in watermarks: 1) distortion-free within a single token generation, 2) distortion-free in one entire generation, 3) distortion-free across multiple generations.

\begin{definition}[Step-wise distortion-free watermark]\label{def:step-wise}
If a watermark framework adopts a distortion-free PDA-rule, then it is a step-wise distortion-free watermark.
\end{definition}
\begin{definition}[Weakly distortion-free watermark]\label{def:weakly}
   A step-wise distortion-free watermark $P_W$ is weakly distortion-free, if $\forall n\in\mathbb{N}_+,\forall\x_{1:n} \in \cV$, we have $P_M(\x_{1:n}) = \bbE_{\k_{1:n}}[P_W(\x_{1:n}|\bm{k}_{1:n})]$.
\end{definition}

\begin{definition}[Strongly distortion-free watermark]\label{def: strongly distortion-free}
    A step-wise distortion-free watermark $P_W$ is strongly distortion-free if for arbitrary number of generation $N_0$ and $\forall\x^{(i)}_{1:n} \in \cV,i\in[N_0]$, it holds that $\prod_{i=1}^{N_0} P_M(\x^{(i)}_{1:n}) = \bbE_{\k^{(1)}_{1:n},...,\k^{(N_0)}_{1:n}}[\prod_{i=1}^{N_0}P_W(\x^{(i)}_{1:n}|\bm{k}^{(i)}_{1:n})]$. 
\end{definition}
In the next theorem, we show the sufficient conditions for achieving a weakly/strongly distortion-free watermark.
 \begin{theorem}\label{thm:sufficient conditions}
     A watermark framework is a weakly/strongly distortion-free watermark if a) it adopts a distortion-free PDA-rule and b) there is no key collision during watermark generation.
 \end{theorem}

 \begin{corollary}
     A watermark that consists of a distortion-free PDA-rule with n-gram hashing, position hashing or fixed key set is a weakly distortion-free watermark.
 \end{corollary}

The proof of this corollary is straightforward because all these watermark key samplers guarantee the uniqueness of each watermark key in a single generation. However, across multiple generations, key collisions become inevitable as the number of generated tokens can surpass the volume of available keys.
  In the rest of this section, we will explain how key collisions can impact the generation quality and lead to a biased watermarked distribution compared to the original language model distribution.

\vspace{-0.2cm}
\subsection{Existing Distortion-Free PDA-Rules}\label{sec:exist pda rule}
\begin{figure}
    \centering
    \includegraphics[width=1.02\textwidth]{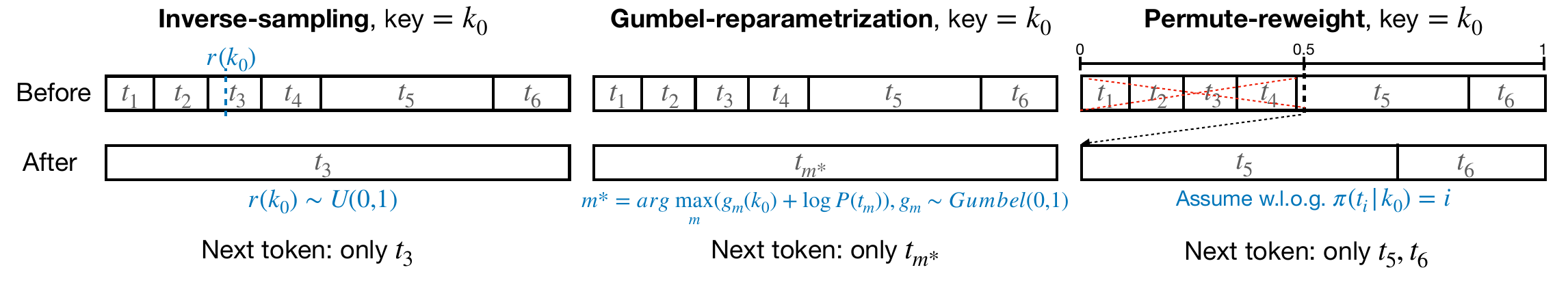}
    \vspace{-0.7cm}
    \caption{Pseudo-randomness in a token sampling step for watermarked LMs. ``Before'' refers the original LM token distribution and ``After'' refers the watermarked token distribution.  Given a fixed watermark key, both inverse-sampling and Gumbel reparametrization methods become deterministic. In contrast, the permute-reweight method retains elements of randomness.}
    \label{fig:comparison}
    \vspace{-0.5cm}
    \label{fig:enter-label}
\end{figure}

\vspace{-0.2cm}
To analyze the influence of key collision on the distortion-free watermarks, we begin with introducing the existing PDA-rules. We also provide a detailed illustration of the existing PDA-rules in Figure~\ref{fig:comparison}.



\textbf{Gumbel-reparametrization.} In the Gumbel-reparametrization rule, when sampling $x_i$ with the watermark key $k_i$, we first sample Gumbel pseudo-random variables $g_1(k_i),...,g_N(k_i)\sim Gumbel(0,1)$ with the watermark key $k_i$. These $N$ independent Gumbel random variables are added to the log-probability of tokens $\log P_M(t_1|\x_{1:i-1}),...,\log P_M(t_N|\x_{1:i-1})$. The token that achieves the maximum value is then selected as the next token $x_i$. This process can be formulated through the following equation: $F_{GR}(P_M(\cdot|\x_{1:i-1}),k_i) = \delta_{t_{m^*}}$, where $m^* = arg\max_{m\in [N]}(g_m(k_i)+\log P_M(t_m|\x_{1:i-1}))$ and $\delta$ is the Dirac function.

\textbf{Inverse-sampling.} In the inverse-sampling rule, when sampling $x_i$ with the watermark key $k_i$, we first organize the LM token probability $ P_M(t_1|\x_{1:i-1}),..., P_M(t_N|\x_{1:i-1})$ within the interval $[0,1]$. Then we will sample a pseudo-random variable $r(k_i)\in U(0,1)$, where $U(0,1)$ is the uniform distribution on $[0,1]$. The next token is selected based on the location of $r(k_i)$ within the cumulative probability intervals on $[0,1]$. This process can be formulated through the following equation:
$F_{IS}(P_M(\cdot|\x_{1:i-1}),k_i) = \delta_{t_{m^*}\in V}$, where $r(k_i) \in [\sum_{j=1}^{m^*-1}P_M(t_j|\x_{1:i-1}), \sum_{j=1}^{m^*}P_M(t_j|\x_{1:i-1})]$ and $\delta$ is the Dirac function.

\textbf{Permute-reweight.} In the permute-reweight rule, when sampling $x_i$ with the watermark key $k_i$, we first generate a pseudo-random token permutation $\pi(\cdot|k_i): V\to [N]$, which is a bijection between token set $V$ and $[N]$. The token permutations are uniformly distributed with the watermark keys. The LM token probabilities are then rearranged within the interval $[0,1]$ according to the permutation $\pi(\cdot|k_i)$. The token probability within $[0,1/2]$ will be scaled to $0$, and the rest half will be scaled to $1$. Subsequently, $x_i$ is randomly sampled following this adjusted distribution. We can formulate the permute-reweight rule through the following formula:
$F_{PR}(P_M(\cdot|\x_{1:i-1}),k_i)(t) = \max\{2\sum_{t',\pi(t'|k_i)\leq\pi(t|k_i)}P_M(t'|\x_{1:i-1})-1,0\}-\max\{2\sum_{t',\pi(t'|k_i)\leq\pi(t|k_i)-1}P_M(t'|\x_{1:i-1})-1,0\}$.
\begin{table*}[t]
\centering
\caption{Summarization of existing distortion-free watermarks.}
\label{tab:watermark comparison}
\resizebox{1\textwidth}{!}{%
\begin{tabular}{l|l|cccc}
\toprule

&& \citet{Aaronson2022}  &\citet{christ2023undetectable} & \citet{kuditipudi2023robust} & \citet{hu2023unbiased}  \\ \midrule
 \multirow{2}{*}{Watermark generator} &PDA-rule   & Gumbel-reparametrization   & Inverse-sampling                 & \begin{tabular}[c]{@{}c@{}}Inverse-sampling, \\ Gumbel-reparametrization\end{tabular}  & \begin{tabular}[c]{@{}c@{}}Inverse-sampling, \\ Permute-reweight\end{tabular}       \\ \cmidrule{2-6}
 &Watermark key sampler   &n-gram hashing                & position hashing              &fixed key set  &n-gram hashing         \\ \midrule
 \multirow{2}{*}{Watermark detector}&Model-agnostic & \Checkmark         & \Checkmark           &\Checkmark &\XSolidBrush           \\
   &Robust    &\Checkmark    & \XSolidBrush          & \Checkmark  &\Checkmark           \\ \midrule
 \multirow{3}{*}{Level of distortion-free} &Step-wise distortion-free &\Checkmark    & \Checkmark          & \Checkmark  &\Checkmark  \\
   &Weakly distortion-free &\Checkmark    & \Checkmark          & \Checkmark  &\Checkmark\\
   &Strongly distortion-free &\XSolidBrush    & \XSolidBrush          & \XSolidBrush  &\XSolidBrush\\
 \bottomrule
\end{tabular}
}
\vspace{-0.3cm}
\end{table*}

\textbf{Pseudo- vs True- Randomness.} 
Based on the above discussion, it is clear that token sampling using Gumbel-reparametrization or inverse-sampling relies entirely on pseudo-randomness, as the watermark distribution for these methods is deterministic given the watermark key. Consequently, for the same token distribution, key collisions result in identical token generation. For instance, when \textbf{generating multiple responses with the same prompt}, the first token will always be identical. In contrast, token sampling with the permute-reweight rule does not fully depend on pseudo-randomness. The permute-reweight PDA-rule only scales the first half of the distribution to zero, preserving the rest of the token probabilities. True-random sampling is then applied to the remaining tokens.


\vspace{-0.2cm}
\subsection{Non-Existence of Strongly Distortion-Free Watermarks under Key Collisions}
\vspace{-0.2cm}

In this subsection, we explore the distribution bias introduced by the watermark.  Given that the distribution overlap between two distributions \(P_1, P_2 \in \mathcal{P}\) is represented by \(\sum_{t \in V} \min\{P_1(t), P_2(t)\}\), we use $1 - \sum_{t\in V}\min\{P_1(t),P_2(t)\}$, i.e., the total variation, to measure the distribution bias between $P_1$ and $P_2$. 
Under the key collisions, the bias introduced by a PDA-rule $F$ on a token distribution $P\in\cP$ is $1 - \sum_{t\in V}\min\{P(t),F(P|k)(t)\}$. Thus, we introduce the \textit{expected total variation} as a metric for measuring distribution bias.



\begin{definition}[Expected total variation]
Given a token distributions $P \in \cP$ and a PDA-rule $F$, the expected total variation between them is given by $\mathbb{D}(P,F) := 1-\bbE_{k}[\sum_{t\in V}\min\{P(t),F(P|k)(t)\}]$.
\end{definition}

\textbf{Trade-off between watermark strength and distribution bias under key collisions.} Interestingly, the expected total variation also reflects the watermark's strength. In statistical watermarking, where the goal is to embed a statistical signal into generated content, a larger total variation enhances the strength of this signal and improve the detection efficiency. 
However, under key collisions, it is desirable for the expected total variation to be minimized to better preserve the original LM distribution. Therefore, a trade-off exists between watermark strength and distribution bias under key collisions.

We compute the expected distribution bias of the existing distortion-free PDA-rules: Gumbel-reparametrization $F_{GR}$, inverse-sampling $F_{IS}$, and permute-reweight $F_{PR}$.
 
\begin{theorem}\label{thm: distribution bias}
Given an arbitrary token distribution $P\in\cP$, we have $\mathbb{D}(P,F_{GR}) =\mathbb{D}(P,F_{IS}) = 1-\sum_{t\in V}P(t)^2,$
and $$0.5(1-\max_{t\in V}P(t))\leq\mathbb{D}(P,F_{PR}) \leq 0.5-\max\{\max_{t\in V}P(t)-0.5,0\}.$$ 
Moreover, $\mathbb{D}(P,F_{PR})\leq \mathbb{D}(P,F_{IS}) = \mathbb{D}(P,F_{GR}).$
\end{theorem}

From this theorem, we find that the permute-reweight watermark exhibits a smaller distribution bias compared to the Gumbel-reparametrization and inverse-sampling watermarks. This finding aligns with our analysis in Section~\ref{sec:exist pda rule}, where we assert that Gumbel-reparametrization and inverse-sampling become deterministic with a fixed watermark key, while permute-reweight maintains an element of randomness, resulting in a smaller distribution bias. In the next theorem, we will show that under key collisions, a watermark with a PDA-rule $F$ is strongly distortion-free if and only if $\mathbb{D}(P,F) = 0,\forall P\in\cP$, which indicates that no signal can be embedded into the generated content.

\begin{theorem}\label{thm:trade-off}
   Under key collisions, a watermark with a distortion-free PDA-rule $F$ is strongly distortion-free if and only if $\forall P\in \cP$, $\mathbb{D}(P,F) = 0$.
\end{theorem}
By integrating Theorem~\ref{thm:trade-off} with Theorem~\ref{thm: distribution bias}, we find that $F_{GR}$, $F_{IS}$, and $F_{PR}$ are unable to yield a strongly distortion-free watermark when key collisions occur. Thus, all existing distortion-free watermarks \citep{Aaronson2022,christ2023undetectable,kuditipudi2023robust,hu2023unbiased} are not strongly distortion-free. Following the above discussion, we summarize the characteristics of existing distortion-free watermarks in Table~\ref{tab:watermark comparison}.
\begin{corollary}
 Under key collisions, a strongly distortion-free watermark does not exist.
\end{corollary} 
 If $\forall P\in \cP,\mathbb{D}(P,F) = 0$, the watermarked LM shows no distribution bias towards the original LM under the watermark key, i.e., $\forall k\in K, F(P|k) = P$. In this case, no watermark is added to the generated content. As key collision is inevitable, we can conclude that with the current watermark key sampling approaches, a strongly distortion-free watermark does not exist.

\vspace{-0.2cm}
\section{Reducing Distribution Bias via Beta-Watermark}
\vspace{-0.2cm}
\begin{wrapfigure}{r}{0.45\textwidth}
    \centering
    \includegraphics[width=0.45\textwidth]{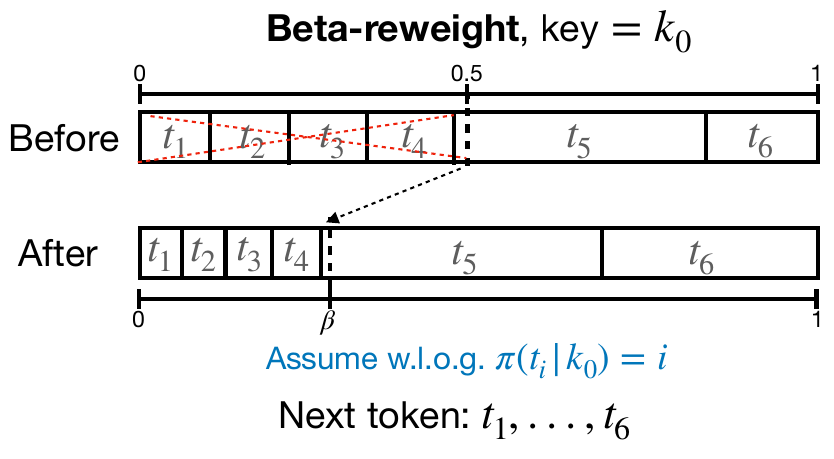}
    \vspace{-0.5cm}
    \caption{Illustration of Beta PDA-rule.}
    \label{fig:beta-watermark}
    \vspace{-0.3cm}
    \label{fig:enter-label}
\end{wrapfigure}
In this section, we focus on reducing the distribution bias resulting from key collisions by introducing a new family of watermarks called beta-watermark. The beta-watermark is based on a distortion-free beta PDA-rule and n-gram hashing. Additionally, we present a novel model-agnostic detection method for it. In Appendix~\ref{sec:algorithms} Alg.~\ref{alg:beta generator} and~\ref{alg:beta detector} we show the algorithms of the generator and detector of beta-watermark.

The beta PDA-rule is a variation of the permute-reweight PDA-rule that introduces greater true randomness during sampling. Similar to permute-reweight watermark, When sampling $x_i$ with the watermark key $k_i$, we first generate a pseudo-random token permutation $\pi(\cdot|k_i): V\to [N]$. Then we arrange the LM token probability within the interval $[0,1]$ following the permutation $\pi(\cdot|k_i)$. The first half of token probability (token probability within $[0,1/2]$) will be scaled to $\beta$, and the rest half probability will be scaled to $1-\beta$ (See Figure~\ref{fig:beta-watermark} for a detailed illustration). The next token is randomly sampled from the new distribution. Notice, when $\beta=0$, the permute-reweight PDA-rule is applied and when $\beta=0.5$, the original LM distribution is used.
\begin{definition}[Beta PDA-rule] Beta PDA-rule $F_{\beta}$ is defined by:
$F_{\beta}(P_M(\cdot|\x_{1:i-1}),k_i)(t) = (1-\beta)F_{PR}(P_M(\cdot|\x_{1:i-1}),k_i)(t)+\beta[\max\{2\sum_{t',\pi(t'|k_i)\geq\pi(t|k_i)}P_M(t'|\x_{1:i-1})-1,0\} - \max\{2\sum_{t',\pi(t'|k_i)\geq\pi(t|k_i)+1}P_M(t'|\x_{1:i-1})-1,0\}].$
Notice, the range of $\beta$ is from $0$ to $0.5$.
\end{definition}

\begin{theorem}\label{thm:beta watermark}
    Beta PDA-rule is a distortion-free PDA-rule, i.e., $\forall \x_{1:n} \in \cV,\forall i \leq n$, $P_M(x_i|\x_{1:i-1}) = \bbE_{k_i}[F(P_M(\cdot|\x_{1:i-1}),k_i)(x_i)]$.
\end{theorem}
\begin{corollary}
    Beta-watermark is a weakly distortion-free watermark.
\end{corollary}
The proof is straightforward, as the beta-watermark consists of a distortion-free PDA-rule and the n-gram hashing. In the subsequent theorem, we theoretically demonstrate that the beta PDA-rule introduces a smaller distribution bias compared to the permute-reweight watermark.
\begin{theorem}\label{thm:beta distribution bias}
    Given an arbitrary token distribution $P\in\cP$, $\mathbb{D}(P,F_\beta)\leq \mathbb{D}(P,F_{PR}) - \beta(1-\max_{t\in V}P(t)).$ Besides, if $\beta_1<\beta_2$, $\mathbb{D}(P,F_{\beta_1})> \mathbb{D}(P,F_{\beta_2})$.
\end{theorem}

As the detector of the permute-reweight watermark \citep{hu2023unbiased} is dependent on the logits from the original LM, we design a new model-agnostic detection algorithm for the beta-watermark. As shown in Figure~\ref{fig:beta-watermark}, beta-reweighting tends to enhance the token probability towards the end of the permutation. During detection, given an input token, we can determine its position within the permutation using $\pi(x|k)$. Thus, a higher score should be assigned to larger values of $\pi(x|k)$. We use a sigmoid function: $\text{sigmoid}(C(\pi(x|k)/|V|-0.5))$, where $C$ is a scaling parameter, to appropriately scale the scores.


\begin{definition}[Model-agnostic beta-reweight detection]\label{def:model agnostic detection} We use score function $s(x,k,F) =sigmoid(C(\pi(x|k)/|V|-0.5))$ to conduct detection. Given a random observation $\x_{1:n}$, under the null hypothesis, we have $\Pr(S(\x_{1:n})-\mathbb{E}_{H_0}[S(\x_{1:n})]>t\sqrt{n}|H_0)\leq\exp(-2t^2)$.
\end{definition}
\vspace{-0.2cm}
\section{Experiments}\label{sec:experiment}
\vspace{-0.2cm}
Our experimental section consists of three parts. 
In the first part, we compare the weakly and strongly distortion-free nature of the beta watermark with that of existing watermarks. In the second part, we evaluate the detection efficiency of the beta watermark against existing watermarks. In the third part, we assess the robustness of the beta watermark when subjected to random paraphrasing attacks. We focus on three seq2seq tasks in our experiments: machine translation, text summarization and text generation. Detailed experimental settings are provided in Appendix~\ref{sec:detailed_experiment_setup} and additional experimental results are in Appendix~\ref{sec:add exp results}.

\vspace{-0.2cm}
\subsection{Distortion-Free}
\vspace{-0.2cm}
In this section, we conduct experiments to validate our theoretical analysis. We evaluate the weakly and strongly distortion-free properties of existing watermark strategies as defined in Definitions~\ref{def:weakly} and~\ref{def: strongly distortion-free}. We validate the weakly distortion-free property by assessing the quality of the watermarked text generated once for each prompt. For the strongly distortion-free property, we examine the quality of the watermarked text for 1000 prompts, where for each prompt we have 100 generations. We define a new metric $\Delta$, which measures the performance gap between the watermarked model and the original LM. For $n$ prompts $p_1,...,p_n$ with $m$ responses for each $g^{p_i}_1,...,g^{p_i}_m$, denoted by $\textrm{Met}$ an arbitrary performance metric (e.g., perplexity), $\Delta\textrm{Met} = \frac{1}{n}\sum_{i=1}^n\frac{1}{m}|\sum_{j=1}^m \textrm{Met}(g^{p_i}_j(\textrm{No watermark}))- \sum_{j=1}^m \textrm{Met}(g^{p_i}_j(\textrm{Watermarked}))|$

\begin{figure}[t]
    \centering
    \includegraphics[width=1\linewidth]{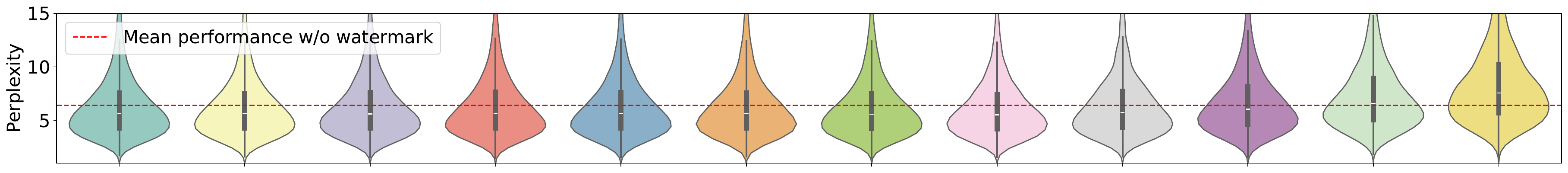}
    \includegraphics[width=1\linewidth]{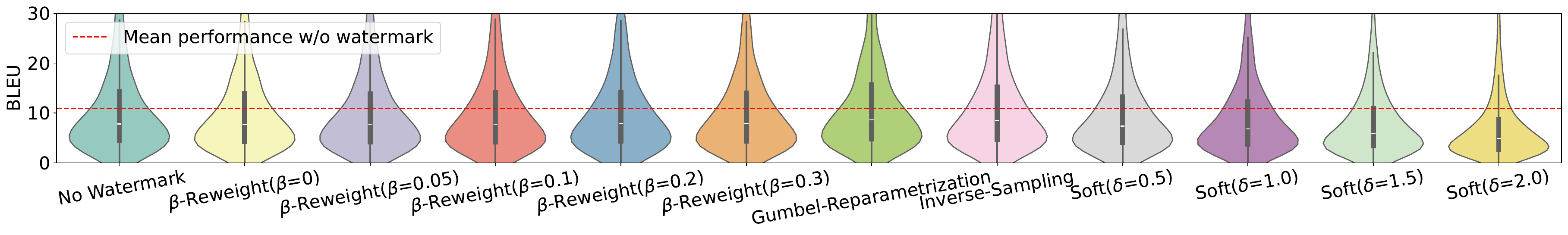}
    \vspace{-0.7cm}
    \caption{Performance of different watermarks under one-time generation.  \textbf{Top:} Violin plot of Text Summarization Perplexity. \textbf{Bottom:} Violin plot of Machine Translation BLEU. We can see the weakly distortion-free watermarks preserve the generation quality.}
    \label{fig:one time generation}
    \vspace{-0.2cm}
\end{figure}

\textbf{Weakly Distortion-Free.}
The results are presented in Figure~\ref{fig:one time generation}. This figure shows that compared to the model without watermarks, all weakly distortion-free watermarks exhibit no significant performance bias in text summarization and text generation tasks. However, for the Soft-watermark~\citep{kirchenbauer2023watermark}, a significant performance bias is observable as $\delta$ increases.


\textbf{Strongly Distortion-Free.}
The results are displayed in Table~\ref{tab:multi-time}. From this table, it is evident that compared to the baseline, which is the $\Delta$ metrics between two non-watermarked models, all weakly distortion-free watermarks demonstrate performance bias across all tasks. In contrast, the Beta-watermark exhibits less bias compared to other weakly distortion-free watermarks. Additionally, as $\beta$ increases, the distribution bias is further reduced, consistent with our theoretical analysis.

\begin{table}[t]
\centering
\caption{Performance of different watermarks under multi-time generations. We randomly selected 1000 prompts and generated 100 responses for each. The baseline is the $\Delta$ metrics between two no-watermarked models.}
\label{tab:multi-time}
\scalebox{0.81}{
\begin{tabular}{l|ccc|cc}
\toprule
 & \multicolumn{3}{c|}{Text Summarization} & \multicolumn{2}{c}{Machine Translation} \\
 & $\Delta$ BERT Score $\downarrow$ & $\Delta$ ROUGE-1$\downarrow$ & $\Delta$ Perplexity $\downarrow$ & $\Delta$ BERT Score $\downarrow$ & $\Delta$ BLEU $\downarrow$ \\ \midrule
Baseline & 0.0062 & 0.0070 & 0.3028 & 0.0180 & 0.7716 \\ \midrule
Beta-Reweight ($\beta=0$) & 0.0090 & 0.0093 & 0.3753 & 0.0267 & 1.2373 \\
Beta-Reweight ($\beta=0.05$) & 0.0084 & 0.0085 & 0.3549 & 0.0248 & 1.1806 \\
Beta-Reweight ($\beta=0.1$) & 0.0079 & 0.0081 & 0.3453 & 0.0230 & 1.0316 \\
Beta-Reweight ($\beta=0.2$) & 0.0070 & 0.0077 & 0.3368 & 0.0203 & 0.9475 \\
Beta-Reweight ($\beta=0.3$) & \textbf{0.0066} &  \textbf{0.0073} &  \textbf{0.3144} &  \textbf{0.0195} &  \textbf{0.8638} \\
Inverse-sampling & 0.0446 & 0.0494 & 1.7846 & 0.1316 & 5.5354 \\
Gumbel-reparametrization & 0.0428 & 0.0488 & 1.8892 & 0.1341 & 5.6438 \\ \midrule
Soft($\delta=0.5$) & 0.0064 & 0.0076 & 0.3331 & 0.0226 & 0.9165 \\
Soft($\delta=1.0$) & 0.0091 & 0.0099 & 0.5473 & 0.0428 & 1.4660 \\
Soft($\delta=1.5$) & 0.0128 & 0.0136 & 1.1237 & 0.0808 & 2.5310 \\
Soft($\delta=2.0$) & 0.0195 & 0.0194 & 2.0817 & 0.1274 & 3.7758 \\ \bottomrule
\end{tabular}
}
\vspace{-0.5cm}
\end{table}

\vspace{-0.2cm}
\subsection{Ablation Study}
\textbf{Detect efficiency.}
We compare the detection efficiency of beta-watermark with Soft-watermark on text generation tasks. We set the detecting scaling parameter (Definition~\ref{def:model agnostic detection}) $C=10$.
We choose the threshold $z = 1.073,1.224,1.517,1.859$, which corresponds to the 10\%, 5\%, 1\% and 0.1\% FPR. From Table~\ref{tab:detect efficiency}, we see that the detect efficiency of beta watermark is comparable with the Soft-watermark~\citep{kirchenbauer2023watermark}. We also see that when $\beta$ increases, the detection efficiency decreases, this is because a larger $\beta$ introduces a smaller distribution bias into the watermarked distribution, thus reducing the watermark strength.

 We use the beta-watermark to illustrate the trade-off between watermark strength and distribution bias. As shown in Figure~\ref{fig:trade-off} (left), with increasing values of $\beta$, the distribution bias decreases, but there is also a corresponding decrease in the true positive rate of watermark detection. This indicates that reducing the distribution bias of the watermark compromises its detectability.

\begin{table}[]
\centering
\caption{ Empirical error rates for watermark detection on text generation. Each row is averaged over around 2000 watermarked examples.}
\label{tab:detect efficiency}
\scalebox{0.9}{
\begin{tabular}{l|l|cc|cc|cc|cc}
\toprule
 &  & \multicolumn{2}{c|}{z=1.073} & \multicolumn{2}{c|}{z=1.224} & \multicolumn{2}{c|}{z=1.517} & \multicolumn{2}{c}{z=1.859} \\
 &  & TNR$\uparrow$ & TPR$\uparrow$ & TNR$\uparrow$ & TPR$\uparrow$ & TNR$\uparrow$ & TPR$\uparrow$ & TNR$\uparrow$ & TPR$\uparrow$ \\ \midrule
\multirow{4}{*}{Soft-watermark} & $\delta=0.5$ & 90.00 & 46.05 & 95.00 & 38.78 & 99.00 & 24.41 & 99.90 & 13.04 \\
 & $\delta=1$ & 90.00 & 88.37 & 95.00 & 85.02 & 99.00 & 76.80 & 99.90 & 68.42 \\
 & $\delta=1.5$ & 90.00 & 97.15 & 95.00 & 96.65 & 99.00 & 94.64 & 99.90 & 90.90 \\
 & $\delta=2$ & 90.00 & 99.45 & 95.00 & 99.39 & 99.00 & 99.06 & 99.90 & 97.90 \\ \midrule
\multirow{5}{*}{Beta-watermark} & $\beta=0$ & 90.00 & 97.75 & 95.00 & 97.17 & 99.00 & 94.69 & 99.90 & 90.25 \\
 & $\beta=0.05$ & 90.00 & 96.82 & 95.00 & 96.19 & 99.00 & 92.67 & 99.90 & 86.26 \\
 & $\beta=0.1$ & 90.00 & 95.76 & 95.00 & 94.19 & 99.00 & 89.13 & 99.90 & 79.90 \\
 & $\beta=0.2$ & 90.00 & 86.53 & 95.00 & 82.49 & 99.00 & 71.14 & 99.90 & 58.55 \\
 & $\beta=0.3$ & 90.00 & 64.59 & 95.00 & 56.88 & 99.00 & 40.67 & 99.90 & 25.38 \\ \bottomrule
\end{tabular}
}
\end{table}





\begin{figure}\label{fig:trade-off}
    \centering
    \includegraphics[width=0.495\textwidth]{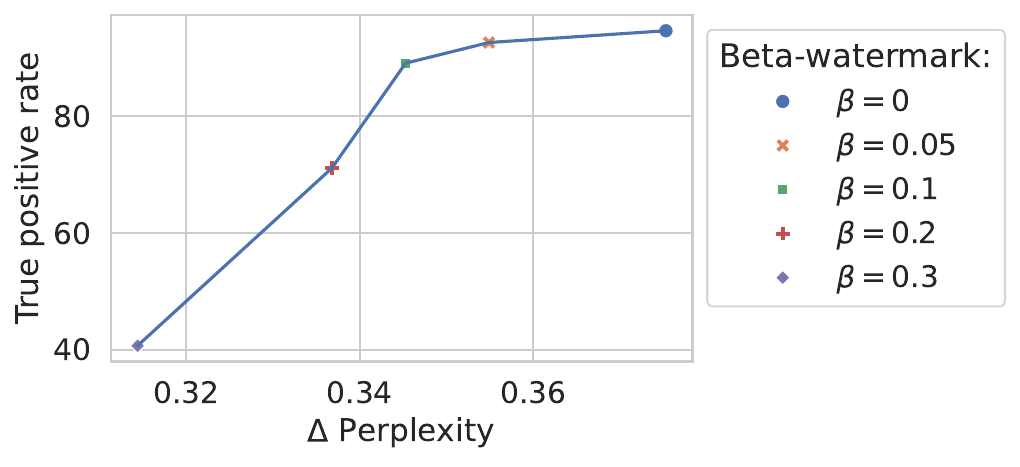}
    \includegraphics[width=0.495\textwidth]{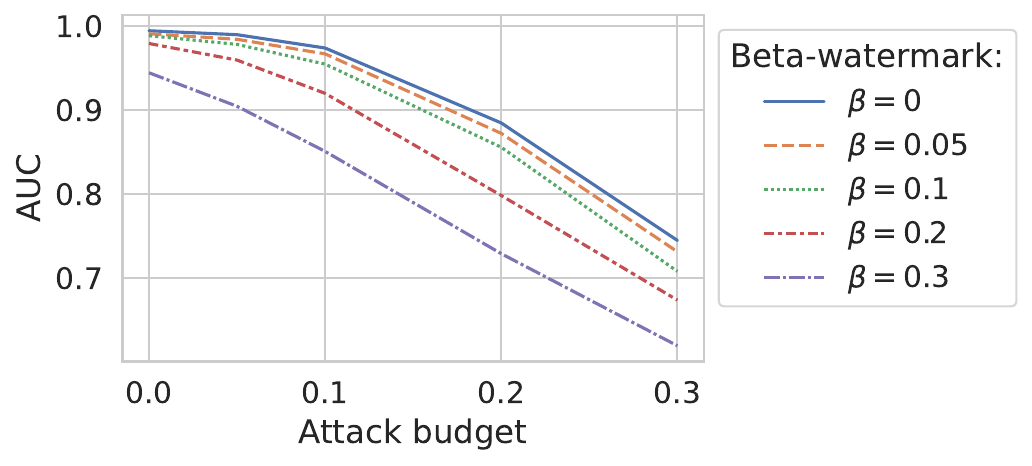}
    \vspace{-0.6cm}
    \caption{\textbf{Left.} Trade-off between distribution bias and watermark strength under key collision. The TPR is measured under 1\% FPR. We can see $\Delta$ Perplexity (distribution bias) increase with the TPR. \textbf{Right.} AUC score of different watermarks under varying attack strength $\epsilon$ on text generation task.}
    \label{fig:robust}
    \vspace{-0.5cm}
\end{figure}

\textbf{Robustness.} We assessed the robustness of the beta-watermark against random text paraphrase attacks \citep{kirchenbauer2023watermark}, where we modified 5\%, 10\%, 20\%, and 30\% (i.e., $\epsilon = 0.05,0.1,0.2,0.3$) of the tokens. The results, as detailed in Figure~\ref{fig:trade-off} (right), indicate that the beta-watermark maintains its robustness even with text modifications up to 30\%.


\vspace{-0.2cm}
\section{Conclusion}\label{sec:conclusion}
\vspace{-0.2cm}
\textbf{Discussion and Limitations.} While we have demonstrated that strongly distortion-free watermarks are not feasible under conditions of watermark key collisions, it remains possible to devise a new watermark key sampling method with an infinite key volume that avoids such collisions. In such a scenario, achieving a strongly distortion-free watermark could still be possible.

In this paper, we demonstrate through extensive theoretical and empirical analysis that existing distortion-free watermarks do not truly preserve the original LM distribution when key collisions occur. Besides, we introduce the beta-watermark to reduce the distribution bias between the watermarked and the original LM distributions. Despite these advances, achieving a strongly distortion-free watermark remains complex, requiring further innovation in watermark key sampling methods. Future research should explore these avenues to enhance the effectiveness and reliability of watermarking techniques in LM.






\bibliography{Styles/neurips_2024}
\bibliographystyle{neurips_2024}
\clearpage
\newpage

\appendix

\section{Algorithms of Beta-watermark}\label{sec:algorithms}
\begin{algorithm}[h]
\caption{Beta-watermark generator}\label{alg:beta generator}
\begin{algorithmic}[1]
\State \textbf{Input:}
secret key $\sk$, parameter $\beta$, prompt $\bm{x}_{-m:0}$, generate length $n\in\mathbb{N}$, texture key history $hist$, n-gram parameter $a$, and permutation generation function $h$.
\For{$i=1,\dots,n$}
        \State Calculate the LM distribution for generating the $i$-th token $P_M(\cdot\mid\bm{x}_{-m:i-1})$. 
        \State Generate a watermark key $k_{i} = (\sk,\x_{i-a,i-1})$.
    \If{$\k_{i} \in hist$}
        \State Sample the next token $x_{i}$ using original LM distribution $P_M(\cdot\mid\bm{x}_{-m:i-1})$.
    \Else
        \State Generate the permutation of token set $\pi(\cdot|k_i)$.
        \State Calculate watermarked distribution $P_W(\cdot|\x_{-m:i-1})=F_{\beta}(P_M(\cdot\mid\bm{x}_{-m:i-1}),k_i)$.
    \State Sample the next token $x_{i}$ using distribution $P_W(\cdot|\x_{-m:i-1})$.
    \EndIf
        
\EndFor
\State \textbf{return} $\bm{x}_{1:n}$.
\end{algorithmic}
\end{algorithm}

\begin{algorithm}[h]
\caption{Beta-watermark detector}\label{alg:beta detector}
\begin{algorithmic}[1]
\State \textbf{Input:}
text $\x_{1:n}$, secret key $\sk$, volume of the token set $N$, score function $s$,  n-gram parameter $a$, threshold $z$.
    \State Initialize the score function: $S=0$.
    \For{$i = 2,...,n$}
    \State Generate the watermark key $k_{i} = (\sk,\x_{i-a,i-1})$.
    \State Generate the permutation of token set $\pi(\cdot|k_i)$.
    \State Update the score function via $S = S + s(\pi(x_i|k_i),k_i,F_{\beta})$.
    \EndFor
\State \textbf{return} $S>z\sqrt{n}$.
\end{algorithmic}
\end{algorithm}
\section{Related Work}\label{sec:add related work}
\textbf{Statistical watermarks.} \cite{kirchenbauer2023watermark} enhanced the statistical watermark framework originally introduced by \cite{Aaronson2022}, demonstrating the effectiveness of statistical watermarking through extensive experiments on large language models. 
They splited the LM tokens into red and green list, then promoted the use of green tokens by adding a fixed parameter $\delta$ to their logits. \cite{zhao2023provable} proposed the unigram watermark, which enhances the robustness of the statistical watermark by using one-gram hashing to produce watermark keys. \cite{liu2023semantic} also improved the robustness of statistical watermarking by leveraging the semantics of generated content as watermark keys. Additionally, \cite{liu2023unforgeable} proposed an unforgeable watermark scheme that employs neural networks to modify token distributions instead of using traditional watermark keys. However, these approaches may lead to significant changes in the distribution of generated text, potentially compromising content quality.

\textbf{Distortion-free watermarks.} To preserve the original output distribution in watermarked content, researchers have explored alternative strategies to modify the token distribution. \cite{Aaronson2022} introduced the first distortion-free watermarking strategy, which utilized Gumbel-reparametrization to alter token distribution and the prefix n-gram content as the watermark keys. \cite{christ2023undetectable} and \cite{kuditipudi2023robust} adopted the inverse-sampling and Gumbel-reparametrization to modify the watermarked token distributions, where the watermark keys is based on the token position or a fixed key list respectively. Notice \cite{christ2023undetectable}'s method encounters resilience challenges under modifications and lacks empirical evidence regarding its detectability. Meanwhile, \cite{kuditipudi2023robust}'s detection process involves hundreds of resampling steps from the secret key distribution, proving inefficient for processing lengthy texts. \cite{hu2023unbiased} employed inverse-sampling and permute-reweight methods for watermarking. But their detector is not model-agnostic, which requires access to the language model API and prompts, which compromises its operational efficiency.

\textbf{Post-hoc Detectors.}
Post-hoc detection serves as a significant alternative to watermarking, focusing on the retrospective analysis of machine-generated text. This can be achieved by leveraging inherent features of language models or by enhancing pre-existing expansive models to function as detectors, as detailed by \citep{zellers2019defending}. Specific implementation nuances, such as sampling methods, can be uncovered through reverse engineering the generated text, a process described by \citep{tay2020reverse}. Additionally, there are post-hoc detectors designed for modern large language models \citep{mitchell2023detectgpt,tian2023gptzero,kirchner2023new}, specifically trained for binary detection tasks. However, there is a growing consensus that these detection methods are becoming less effective as language models evolve. As observed by \cite{gambini2022pushing}, detection mechanisms effective against GPT-2 have struggled with GPT-3. Moreover, text rephrasing models like those in \citep{krishna2023paraphrasing} are bypassing prevalent post-hoc detectors such as GPTZero \citep{tian2023gptzero}, DetectGPT \citep{mitchell2023detectgpt}, and OpenAI's proprietary detector \citep{kirchner2023new}. Additionally, \cite{chakraborty2023possibilities} notes that as AI-generated content becomes increasingly indistinguishable from human-produced text, the demand on post-hoc detectors to analyze longer text segments will likely increase.

\textbf{Steganography.}
Steganography involves embedding hidden messages in media such as natural language or images, ensuring only intended recipients can detect the message while it remains concealed from others \citep{hopper2002provably}. When applied to watermarking, the goal is to maintain stealth. However, established steganography techniques may not meet this goal without certain entropy-related assumptions. In scenarios where language model prompts can be adversarially chosen, the need for stealth remains. This discrepancy arises due to the different levels of access that watermarking and steganography have to the model's output distribution. In steganography, there is only oracle access to this distribution, whereas our watermarking approach provides a detailed view of the token's probability distribution. Hence, while steganography either depends on entropy assumptions \citep{hopper2002provably} or risks security with low entropy channels \citep{dedic2009upper}, our watermark remains stealthy regardless of the text's entropy. This is achieved by leveraging full distribution access and using it as a foundation for embedding watermarks. \cite{kaptchuk2021meteor} discusses encoding with similar access but presupposes equal decoding access, which is impractical for watermarking as the detection algorithm typically lacks the initiating prompt, thus remaining unaware of the distribution.

\section{Missing Proofs}\label{sec:missing proof}
\subsection{Proof of Theorem~\ref{thm:multikey detection}}
\begin{proof}
    \begin{equation}
        \begin{split}
            \Pr\left(\max_{i\in[M]}(S(\x_{1:n}|\sk_i)-\bbE_{H_0}[S])\leq t|H_0\right) &= \prod_{i=1}^M\Pr\left(S(\x_{1:n}|\sk_i)-\bbE_{H_0}[S]\leq t|H_0\right)\\
             &= \prod_{i=1}^M(1-\Pr\left(S(\x_{1:n}|\sk_i)-\bbE_{H_0}[S]\geq t|H_0\right)) \\
            &= (1-p_0(t))^M.
        \end{split}
    \end{equation}
Thus, $\Pr\left(\max_{i\in[M]}(S(\x_{1:n}|\sk_i)-\bbE_{H_0}[S])\geq t|H_0\right) = 1-\Pr\left(\max_{i\in[M]}(S(\x_{1:n}|\sk_i)-\bbE_{H_0}[S])\leq t|H_0\right)= 1-(1-p_0(t))^M. $
\end{proof}
\subsection{Proof of Theorem~\ref{thm:sufficient conditions}}
 \begin{proof}
We first show the weakly distortion-free case: firstly, if key collision does not occur, we have
\begin{equation}
    \begin{split}
        \bbE_{\k_{1:n}}[P_W(\x_{1:n}|\k_{1:n})] &= \bbE_{\k_{1:n}}\left[\prod_{i=1}^nF(P_M(x_i|\x_{1:i-1}),k_i)\right] \\
        &= \prod_{i=1}^n\bbE_{k_i}[F(P_M(x_i|\x_{1:i-1}),k_i)].
    \end{split}
\end{equation}
The above equality stems from the independence property of the PDA-rule $F(P_M(x_i|\x_{1:i-1}),k_i)$. \cite{christ2023undetectable} and \cite{hu2023unbiased} show that if there is no repeating keys in $\k_{1:n}$, the independence property can be satisfied with hash functions. 

Since $F$ is a distortion-free PDA-rule, we have $\bbE_{k_i}[F(P_M(x_i|\x_{1:i-1}),k_i)] = P_M(x_i|\x_{1:i-1})$. Thus, 

\begin{equation}
    \begin{split}
        \bbE_{\k_{1:n}}[P_W(\x_{1:n}|\k_{1:n})] 
        &= \prod_{i=1}^n\bbE_{k_i}[F(P_M(x_i|\x_{1:i-1}),k_i)]=
        \prod_{i=1}^nP_M(x_i|\x_{1:i-1}) = P_M(\x_{1:n}).
    \end{split}
\end{equation}

Analogously, for the strongly distortion-free case, if key collision does not occur, we will have distinct $\k_{1:n}^{(i)}$. By the independence property of the PDA-rule, we have
\begin{equation}
    \begin{split}
    \bbE_{\k^{(1)}_{1:n},...,\k^{(N_0)}_{1:n}}[\prod_{i=1}^{N_0}P_W(\x^{(i)}_{1:n}|\bm{k}^{(i)}_{1:n})] &= \prod_{i=1}^{N_0}\bbE_{\k^{(i)}_{1:n}}[P_W(\x^{(i)}_{1:n}|\bm{k}^{(i)}_{1:n})]\\ 
    &= \prod_{i=1}^{N_0}\prod_{j=1}^{n}\bbE_{k^{(i)}_{j}}[P_W(x^{(i)}_{j}|\x^{(i)}_{1:j-1},k^{(i)}_{j})]\\
    & = \prod_{i=1}^{N_0}\prod_{j=1}^{n}\bbE_{k^{(i)}_{j}}[F(P_M(x^{(i)}_{j}|\x^{(i)}_{1:j-1}),k^{(i)}_{j})]\\
    & = \prod_{i=1}^{N_0}\prod_{j=1}^{n}P_M(x^{(i)}_{j}|\x^{(i)}_{0:j-1})\\
    & = \prod_{i=1}^{N_0}P_M(\x^{(i)}_{1:n}).
    \end{split}
\end{equation}

\end{proof}

\subsection{Proof of Theorem~\ref{thm: distribution bias}}\label{proof:distribution bias}
\begin{proof}
\textbf{Part 1.} We start from proving $\mathbb{D}(P,F_{GR}) =\mathbb{D}(P,F_{IS}) = 1-\sum_{t\in V}P(t)^2$. Since both $F_{GR}$ and $F_{IS}$ are distortion-free PDA-rule, $P(t) = \bbE_k[F_{GR}(P|k)(t)] = \bbE_k[F_{IS}(P|k)(t)]$. Since $F_{GR}(P|k)$ and $F_{IS}(P|k)$ are Dirac distribution, when $F_{GR}(P|k)(t)>0$, $F_{GR}(P|k)(t)=1$, and $\bbE_k[F_{GR}(P|k)(t)] =\bbE_k[\bm{1}_{F_{GR}(P|k)(t)>0}]=\Pr(F_{GR}(P|k)(t)>0),\forall t\in V$.  Thus,
\begin{equation}
    \begin{split}
        \bbE_{k}[\sum_{t\in V}\min\{P(t),F_{GR}(P|k)(t)\}] &= \sum_{t\in V}\bbE_{k}[P(t)\bm{1}_{F_{GR}(P|k)(t)>0}]\\
        & = \sum_{t\in V}\bbE_{k}[P(t)\bm{1}_{F_{GR}(P|k)(t)>0}]\\
        & = \sum_{t\in V}\bbE_{k}[P(t)|\bm{1}_{F_{GR}(P|k)(t)>0}]\Pr(F_{GR}(P|k)(t)>0)\\
        & = \sum_{t\in V} P(t)^2
    \end{split}
\end{equation}
Analogously, $\bbE_{k}[\sum_{t\in V}\min\{P(t),F_{IS}(P|k)(t)\}]=\sum_{t\in V} P(t)^2$. Therefore, we have 
\begin{equation}\label{eqn:gris}
    \mathbb{D}(P,F_{GR}) =\mathbb{D}(P,F_{IS}) = 1-\sum_{t\in V}P(t)^2
\end{equation}.

\textbf{Part 2.} Next, we show $0.5(1-\max_{t\in V}P(t))\leq\mathbb{D}(P,F_{PR})\leq 0.5 - \max\{\max_{t\in V}P(t)-0.5,0\}$. Given a permutation on the token list, assume w.l.o.g. the permutation is of order $\{t_1,...,t_N\}$, in $F_{PR}$ we will arrange the token probabilities on the interval $[0,1]$ following the permutation order. Denote by $i_0$ the index of the token such that 0.5 lies in its probability region, then the token probabilities of $\{t_{i_0+1},...t_{N}\}$ will be doubled, while the token probabilities of $\{t_{1},...t_{i_0-1}\}$ will be scaled to 0. Thus, under this permutation,
$$\sum_{t\in V}\min\{P(t),F_{PR}(P|k)(t)\} = \sum_{i=i_0+1}^NP(t_i)+\min\{P(t_{i_0}),2\xi_{i_0}\},$$
where $\xi_{i_0}$ is the probability mass of $t_{i_0}$ that is in the interval $[0.5,1]$, $\max\{P(t_{i_0})-0.5,0\}\leq\xi_{i_0}\leq\min\{0.5,P(t_{i_0})\}$. Next, we consider the reverse permutation $\{t_N,...,t_1\}$, following the similar discussion, we have $$\sum_{t\in V}\min\{P(t),F_{PR}(P|k^r)(t)\} = \sum_{i=1}^{i_0-1}P(t_i)+\min\{P(t_{i_0}),2(P(t_{i_0})-\xi_{i_0})\},$$
where $k^r$ refers the key that lead to the reserved permutation. Thus,
\begin{equation}\label{eqn:PR1}
    \begin{split}
        &\sum_{t\in V}\min\{P(t),F_{PR}(P|k)(t)\}+\sum_{t\in V}\min\{P(t),F_{PR}(P|k^r)(t)\}\\
        =&1+\min\{P(t_{i_0}),2\xi_{i_0}\}+\min\{P(t_{i_0}),2(P(t_{i_0})-\xi_{i_0})\}-P(t_{i_0}).\\
    \end{split}
\end{equation}
Next, we show $P(t_{i_0})\geq\min\{P(t_{i_0}),2\xi_{i_0}\}+\min\{P(t_{i_0}),2(P(t_{i_0})-\xi_{i_0})\}-P(t_{i_0})\geq \max\{\max_{t\in V}P(t)-0.5,0\}$. The left hand side inequality is trivial, as $\min\{P(t_{i_0}),2\xi_{i_0}\}+\min\{P(t_{i_0}),2(P(t_{i_0})-\xi_{i_0})\}\leq 2P(t_{i_0})$. 

For the right hand side inequality, given $\min\{A,2x\}+\min\{A,2A-2x\} = A+\min\{2A-2x,2x\}$, we have 
\begin{equation}
    \begin{split}\min\{P(t_{i_0}),2\xi_{i_0}\}+\min\{P(t_{i_0}),2(P(t_{i_0})-\xi_{i_0})\}-P(t_{i_0}) = 2\min\{P(t_{i_0})-\xi_{i_0},\xi_{i_0}\}.
    \end{split}
\end{equation}
Since $0\leq\max\{P(t_{i_0})-0.5,0\}\leq\xi_{i_0}\leq\min\{0.5,P(t_{i_0})\}\leq P(t_{i_0})$, the minimum value of $\min\{P(t_{i_0})-\xi_{i_0},\xi_{i_0}\}$ when $\xi_{i_0}$ take either $\max\{P(t_{i_0})-0.5,0\}$ or $\min\{0.5,P(t_{i_0})\}$, thus
\begin{equation}
    \begin{split}
        \min\{P(t_{i_0})-\xi_{i_0},\xi_{i_0}\}\geq \max\{P(t_{i_0})-0.5,0\}.
    \end{split}
\end{equation}
If $P(t_{i_0})- 0.5>0$, it is obviously $\max_{t\in V}P(t) = P(t_{i_0})$, so 
\begin{equation}
    \begin{split}
        \min\{P(t_{i_0})-\xi_{i_0},\xi_{i_0}\}\geq \max\{\max_{t\in V}P(t)-0.5,0\}.
    \end{split}
\end{equation}

Combining it with Equation~\ref{eqn:PR1}, we have 
\begin{equation}\label{eqn:PRR1}
    \begin{split}
        &\sum_{t\in V}\min\{P(t),F_{PR}(P|k)(t)\}+\sum_{t\in V}\min\{P(t),F_{PR}(P|k^r)(t)\}\\
        =&1+\min\{P(t_{i_0}),2\xi_{i_0}\}+\min\{P(t_{i_0}),2(P(t_{i_0})-\xi_{i_0})\}-P(t_{i_0}).\\
        \leq& 1+P(t_{i_0})\leq 1+\max_{t\in V}P(t),
    \end{split}
\end{equation}
and
\begin{equation}\label{eqn:PRR2}
    \begin{split}
        &\sum_{t\in V}\min\{P(t),F_{PR}(P|k)(t)\}+\sum_{t\in V}\min\{P(t),F_{PR}(P|k^r)(t)\}\\
        =&1+\min\{P(t_{i_0}),2\xi_{i_0}\}+\min\{P(t_{i_0}),2(P(t_{i_0})-\xi_{i_0})\}-P(t_{i_0}).\\
        &\geq 1+2\max\{\max_{t\in V}P(t)-0.5,0\}.
    \end{split}
\end{equation}
Since the permutation over $V$ is uniformly seeded with the watermark keys,
\begin{equation}\label{eqn:PR3}
    \begin{split}
    \mathbb{D}(P,F_{PR}) &= 1-\bbE_{k}[\sum_{t\in V}\min\{P(t),F_{PR}(P|k)(t)\}]\\
    & = 1-\frac{1}{2}\bbE_{k}[\sum_{t\in V}\min\{P(t),F_{PR}(P|k)(t)\}+\sum_{t\in V}\min\{P(t),F_{PR}(P|k^r)(t)\}]
    \end{split}
\end{equation}
Combining it with Equation~\ref{eqn:PRR1} and Equation~\ref{eqn:PRR2}, we have
\begin{equation}
    \begin{split}
    0.5(1-\max_{t\in V}P(t))\leq\mathbb{D}(P,F_{PR})  \leq 0.5 - \max\{\max_{t\in V}P(t)-0.5,0\}.
    \end{split}
\end{equation}

\textbf{Part 3.} Finally, we show $\mathbb{D}(P,F_{PR})\leq \mathbb{D}(P,F_{IS}) = \mathbb{D}(P,F_{GR})$. We only need to prove $ 0.5 - \max\{\max_{t\in V}P(t)-0.5,0\}\leq 1-\sum_{t\in V}P(t)^2$. We have two steps for Part 3.

\begin{lemma}Given $0\leq x_1,x_2\leq r_0\leq r_1$, $x_1+x_2 = r_1\leq 1$, we have $x_1^2+x_2^2\leq r_0^2+(r_1-r_0)^2$.
\end{lemma}
\textbf{Proof.}
    $x_1^2+x_2^2 = x_1^2+(r_1-x_1)^2 = 2x_1^2-2x_1r_1+r_1^2 = 2(x_1-r_1/2)^2+r_1^2/2\leq \min_{x_1}2(x_1-r_1/2)^2+r_1^2/2 =  r_0^2+(r_1-r_0)^2$.

Thus, by inductive we have $1-\sum_{t\in V}P(t)^2\geq 1-\lfloor\frac{1}{\max_{t\in V}P(t)}\rfloor(\max_{t\in V}P(t))^2 - (1-\lfloor\frac{1}{\max_{t\in V}P(t)}\rfloor\max_{t\in V}P(t))^2$.
Now we continue the proof of the main theorem.

\textbf{Step 1.} When $\max_{t\in V}P(t)\geq 0.5$, 
\begin{equation}\label{eqn:PR4}
    \begin{split}
1-\sum_{t\in V}P(t)^2&\geq 1-(\max_{t\in V}P(t))^2 - (1-\max_{t\in V}P(t))^2\\
& = 2\max_{t\in V}P(t) - 2(\max_{t\in V}P(t))^2\\
& = 0.5-2(\max_{t\in V}P(t) - 0.5)^2\\
&\geq 0.5 - (\max_{t\in V}P(t) - 0.5)\\
& = 0.5 - \max\{\max_{t\in V}P(t)-0.5,0\}
    \end{split}
\end{equation}

\textbf{Step 2.} When $\max_{t\in V}P(t)\leq 0.5$, 
\begin{equation}\label{eqn:PR5}
    \begin{split}
1-\sum_{t\in V}P(t)^2&\geq 1-\lfloor\frac{1}{\max_{t\in V}P(t)}\rfloor(\max_{t\in V}P(t))^2 - (1-\lfloor\frac{1}{\max_{t\in V}P(t)}\rfloor\max_{t\in V}P(t))^2\\
& = 2\lfloor\frac{1}{\max_{t\in V}P(t)}\rfloor\max_{t\in V}P(t) - (\lfloor\frac{1}{\max_{t\in V}P(t)}\rfloor+\lfloor\frac{1}{\max_{t\in V}P(t)}\rfloor^2)(\max_{t\in V}P(t))^2,\\
    \end{split}
\end{equation}
denote by $\epsilon = \frac{1}{\max_{t\in V}P(t)} - \lfloor\frac{1}{\max_{t\in V}P(t)}\rfloor$, we have $0\leq\epsilon<1$ and 
\begin{equation}\label{eqn:PR6}
    \begin{split}
1-\sum_{t\in V}P(t)^2
& = 2(\frac{1}{\max_{t\in V}P(t)}-\epsilon)\max_{t\in V}P(t) - ((\frac{1}{\max_{t\in V}P(t)}-\epsilon)+(\frac{1}{\max_{t\in V}P(t)}-\epsilon)^2)(\max_{t\in V}P(t))^2,\\
& = 2-2\epsilon\max_{t\in V}P(t) - \left(\max_{t\in V}P(t)-\epsilon\max_{t\in V}P(t)^2 + 1-2\epsilon\max_{t\in V}P(t)+\epsilon^2\max_{t\in V}P(t)^2\right)\\
& = 1-\max_{t\in V}P(t)+(\epsilon-\epsilon^2)\max_{t\in V}P(t)^2\\
&\geq 1-\max_{t\in V}P(t)\geq 0.5 = 0.5 - \max\{\max_{t\in V}P(t)-0.5,0\}.
    \end{split}
\end{equation}
By Step 1 and Step 2, we have $\mathbb{D}(P,F_{PR})\leq \mathbb{D}(P,F_{IS}) = \mathbb{D}(P,F_{GR})$.
\end{proof}

\subsection{Proof of Theorem~\ref{thm:trade-off}}
\begin{proof}
    Consider the scenario of generating multiple responses with the \textbf{same-prompt single-token-generation} task. According to Definition~\ref{def: strongly distortion-free}
under the strongly distortion-free condition, one must have $\forall P_M\in\cP,\forall N_0\in\mathbb{N}_+,\forall t^{(i)}\in V$, $\prod_{i=1}^{N_0}P_M(t^{(i)}) = \bbE_{k^{(1)},...,k^{(N_0)}}[\prod_{i=1}^{N_0}F(P_M|k^{(i)})(t^{(i)})]$. Under key collisions, there exists at least two $k^{(i)},k^{(j)}$ are the same. Then we have $\forall P_M\in\cP,\exists N_0\geq 2,\forall t^{(i)}\in V$, $\prod_{i=1}^{N_0}P_M(t^{(i)}) = \bbE_{k}[\prod_{i=1}^{N_0}F(P_M|k)(t^{(i)})]$. We will show that this hold if and only if $\mathbb{D}(P_M,F) = 0$.

\textbf{Part 1.} It is obviously that $\mathbb{D}(P_M,F) = 0$ can lead to $\forall N_0\in\mathbb{N}_+,\forall t^{(i)}\in V$, $\prod_{i=1}^{N_0}P_M(t^{(i)}) = \bbE_{k}[\prod_{i=1}^{N_0}F(P_M|k)(t^{(i)})]$. This is because if $\mathbb{D}(P_M,F) = 0$, $P_M(t^{(i)}) = F(P_M|k)(t^{(i)})$ almost surely and thus $\bbE_{k}[\prod_{i=1}^{N_0}F(P_M|k)(t^{(i)})] = \bbE_{k}[\prod_{i=1}^{N_0}P_M(t^{(i)})]=\prod_{i=1}^{N_0}P_M(t^{(i)})$.

\textbf{Part 2.} Now we will show that if $\exists N_0\geq 2,\forall t^{(i)}\in V$, $\prod_{i=1}^{N_0}P_M(t^{(i)}) = \bbE_{k}[\prod_{i=1}^{N_0}F(P_M|k)(t^{(i)})]$, then $\mathbb{D}(P_M,F) = 0$.

As $t^{(i)}$ is arbitrary selected, we can choose $t^{(1)}=,...,=t^{(N_0)}=t$, then we have $P_M(t)^{N_0} =\bbE_k[F(P_M|k)(t)^{N_0}]$. By Jensen's inequality, when $N_0\geq 2$, 
$$P_M(t)^{N_0} =\bbE_k[F(P_M|k)(t)^{N_0}]\geq (\bbE_k[F(P_M|k)(t)])^{N_0} = P_M(t)^{N_0}.$$ 
The equality is achieved if and only if $F(P_M|k)(t) = \bbE_k[F(P_M|k)(t)] = P_M(t)$. Thus, $\forall t\in V,\forall k\in K, F(P_M|k)(t) = P_M(t)$, which leads to
$$\mathbb{D}(P_M,F)  = 1-\bbE_{k}[\sum_{t\in V}\min\{P_M(t),F(P_M|k)(t)\}] = 1-\sum_{t\in V}P_M(t) = 0.$$
\end{proof}

\subsection{Proof of Theorem~\ref{thm:beta watermark}}
\begin{proof}
    We need to show $P_M(t|\x_{1:i-1}) = \bbE_{k_i}[F_{\beta}(P_M(\cdot|\x_{1:i-1}),k_i)(t)]$. As $F_{PR}(P_M(\cdot|\x_{1:i-1}),k_i)(t)$ is a distortion-free PDA-rule, we know $\bbE_{k_i}[(1-\beta)F_{PR}(P_M(\cdot|\x_{1:i-1}),k_i)(t)] = (1-\beta)P_M(t|\x_{1:i-1})$. Thus, we need to show 

    \begin{equation}
    \begin{split}&\bbE_{k_i}\left[\max\{2\sum_{t',\pi(t'|k_i)\geq\pi(t|k_i)}P_M(t'|\x_{1:i-1})-1,0\} - \max\{2\sum_{t',\pi(t'|k_i)\geq\pi(t|k_i)+1}P_M(t'|\x_{1:i-1})-1,0\}\right] \\ &= P_M(t|\x_{1:i-1})
    \end{split}
    \end{equation}
    
    Since the permutation is uniformly distributed, denoted by $\Pi$ the set of all permutations on $V$ and $P_{\Pi}$ the uniformly distribution on $\Pi$, we have
\begin{equation}
    \begin{split}&\bbE_{k_i}\left[\max\{2\sum_{t',\pi(t'|k_i)\geq\pi(t|k_i)}P_M(t'|\x_{1:i-1})-1,0\} - \max\{2\sum_{t',\pi(t'|k_i)\geq\pi(t|k_i)+1}P_M(t'|\x_{1:i-1})-1,0\}\right] \\ 
    = &\bbE_{\pi\sim P_{\Pi}}\left[\max\{2\sum_{t',\pi(t'|k_i)\geq\pi(t|k_i)}P_M(t'|\x_{1:i-1})-1,0\} - \max\{2\sum_{t',\pi(t'|k_i)\geq\pi(t|k_i)+1}P_M(t'|\x_{1:i-1})-1,0\}\right] \\ 
    \end{split}
    \end{equation}
    As $P_{\Pi}$ is the uniformly distribution on $\Pi$, for each $\pi\in\Pi$, we consider its reverse permutation $\pi^r$:
    \begin{equation}
    \begin{split}
    &\bbE_{\pi\sim P_{\Pi}}\left[\max\{2\sum_{t',\pi(t'|k_i)\geq\pi(t|k_i)}P_M(t'|\x_{1:i-1})-1,0\} - \max\{2\sum_{t',\pi(t'|k_i)\geq\pi(t|k_i)+1}P_M(t'|\x_{1:i-1})-1,0\}\right] \\ 
    = &\frac{1}{2}\bbE_{\pi^r\sim P_{\Pi}}\left[\max\{2\sum_{t',\pi(t'|k_i)\geq\pi(t|k_i)}P_M(t'|\x_{1:i-1})-1,0\} - \max\{2\sum_{t',\pi(t'|k_i)\geq\pi(t|k_i)+1}P_M(t'|\x_{1:i-1})-1,0\}\right.\\
    +& \left.\max\{2\sum_{t',\pi^r(t'|k_i)\geq\pi^r(t|k_i)}P_M(t'|\x_{1:i-1})-1,0\} - \max\{2\sum_{t',\pi^r(t'|k_i)\geq\pi^r(t|k_i)+1}P_M(t'|\x_{1:i-1})-1,0\}
    \right]
    \end{split}
    \end{equation}
Notice, if $\pi(t')\leq\pi(t)$, then in the reversed permutation $\pi^r$, we have $\pi^r(t')\geq\pi^r(t)$ and vice versa. Thus,
\begin{equation}
    \begin{split}
    & \max\{2\sum_{t',\pi^r(t'|k_i)\geq\pi^r(t|k_i)}P_M(t'|\x_{1:i-1})-1,0\} - \max\{2\sum_{t',\pi^r(t'|k_i)\geq\pi^r(t|k_i)+1}P_M(t'|\x_{1:i-1})-1,0\}\\
    =&\max\{2\sum_{t',\pi(t'|k_i)\leq\pi(t|k_i)}P_M(t'|\x_{1:i-1})-1,0\} - \max\{2\sum_{t',\pi(t'|k_i)\leq\pi(t|k_i)-1}P_M(t'|\x_{1:i-1})-1,0\}\\
    =&\max\{1-2\sum_{t',\pi(t'|k_i)\geq\pi(t|k_i)+1}P_M(t'|\x_{1:i-1}),0\} - \max\{1-2\sum_{t',\pi(t'|k_i)\geq\pi(t|k_i)-1}P_M(t'|\x_{1:i-1}),0\}.
    \end{split}
\end{equation}
By $\max\{x,0\}-\max\{-x,0\} = x$ we have
\begin{equation}
    \begin{split}
    &\bbE_{\pi\sim P_{\Pi}}\left[\max\{2\sum_{t',\pi(t'|k_i)\geq\pi(t|k_i)}P_M(t'|\x_{1:i-1})-1,0\} - \max\{2\sum_{t',\pi(t'|k_i)\geq\pi(t|k_i)+1}P_M(t'|\x_{1:i-1})-1,0\}\right] \\ 
    = &\frac{1}{2}\bbE_{\pi\sim P_{\Pi}}\left[\max\{2\sum_{t',\pi(t'|k_i)\geq\pi(t|k_i)}P_M(t'|\x_{1:i-1})-1,0\} - \max\{2\sum_{t',\pi(t'|k_i)\geq\pi(t|k_i)+1}P_M(t'|\x_{1:i-1})-1,0\}\right.\\
    +& \left.\max\{2\sum_{t',\pi^r(t'|k_i)\geq\pi^r(t|k_i)}P_M(t'|\x_{1:i-1})-1,0\} - \max\{2\sum_{t',\pi^r(t'|k_i)\geq\pi^r(t|k_i)+1}P_M(t'|\x_{1:i-1})-1,0\}
    \right]\\
    = &\frac{1}{2}\bbE_{\pi\sim P_{\Pi}}\left[\max\{2\sum_{t',\pi(t'|k_i)\geq\pi(t|k_i)}P_M(t'|\x_{1:i-1})-1,0\} - \max\{2\sum_{t',\pi(t'|k_i)\geq\pi(t|k_i)+1}P_M(t'|\x_{1:i-1})-1,0\}\right.\\
    +& \left.\max\{1-2\sum_{t',\pi(t'|k_i)\geq\pi(t|k_i)+1}P_M(t'|\x_{1:i-1}),0\} - \max\{1-2\sum_{t',\pi(t'|k_i)\geq\pi(t|k_i)-1}P_M(t'|\x_{1:i-1}),0\}
    \right]\\
    =&\frac{1}{2}\bbE_{\pi\sim P_{\Pi}}[2\sum_{t',\pi^r(t'|k_i)\geq\pi^r(t|k_i)}P_M(t'|\x_{1:i-1})-1 - (2\sum_{t',\pi^r(t'|k_i)\geq\pi^r(t|k_i)+1}P_M(t'|\x_{1:i-1})-1)]\\
    =&\frac{1}{2}\bbE_{\pi\sim P_{\Pi}}[2P_M(t|\x_{1:i-1})]\\
    =&P_M(t|\x_{1:i-1}).
    \end{split}
    \end{equation}

\end{proof}

\subsection{Proof of Theorem~\ref{thm:beta distribution bias}}
\begin{proof}
    \textbf{Part 1.} We first show $\forall P\in\cP$, $\mathbb{D}(P,F_\beta)\leq \mathbb{D}(P,F_{PR}) - \beta(1-\max_{t\in V}P(t)).$ According to the Part 2 of Proof~\ref{proof:distribution bias}, we know that given a permutation $\{t_1,...,t_N\}$ and let $t_{i_0}$ is the token whose probability mass expands across $1/2$,
$$\sum_{t\in V}\min\{P(t),F_{PR}(P|k)(t)\} = \sum_{i=i_0+1}^NP(t_i)+\min\{P(t_{i_0}),2\xi_{i_0}\},$$
where $\xi_{i_0}$ is the probability mass of $t_{i_0}$ that is in the interval $[0.5,1]$ (notice $t_{i_0}$ is the same for both permuta-reweight and beta PDA-rule as they use the same permutation), $\max\{P(t_{i_0})-0.5,0\}\leq\xi_{i_0}\leq\min\{0.5,P(t_{i_0})\}$. And $$\sum_{t\in V}\min\{P(t),F_{PR}(P|k^r)(t)\} = \sum_{i=1}^{i_0-1}P(t_i)+\min\{P(t_{i_0}),2(P(t_{i_0})-\xi_{i_0})\},$$
where $k^r$ refers the key that lead to the reserved permutation.

Now we consider $F_\beta$, from the similar analysis we have
$$\sum_{t\in V}\min\{P(t),F_{\beta}(P|k)(t)\} = \sum_{i=i_0+1}^NP(t_i)+2\beta\sum_{i=1}^{i_0-1}P(t_i)+\min\{P(t_{i_0}),2(1-\beta)\xi_{i_0}+2\beta(P(t_{i_0}) - \xi_{i_0})\},$$
and 
$$\sum_{t\in V}\min\{P(t),F_{\beta}(P|k^r)(t)\} = \sum_{i=1}^{i_0-1}P(t_i)+2\beta\sum_{i=i_0+1}^NP(t_i)+\min\{P(t_{i_0}),2(1-\beta)(P(t_{i_0})-\xi_{i_0})+2\beta\xi_{i_0}\}.$$
As\begin{equation}
    \begin{split}
        &\min\{P(t_{i_0}),2(1-\beta)\xi_{i_0}+2\beta(P(t_{i_0}) - \xi_{i_0})\}+\min\{P(t_{i_0}),2(1-\beta)(P(t_{i_0})-\xi_{i_0})+2\beta\xi_{i_0}\}\\
        & = P(t_{i_0}) + \min\{2(1-\beta)(P(t_{i_0})-\xi_{i_0})+2\beta\xi_{i_0},2(1-\beta)\xi_{i_0}+2\beta(P(t_{i_0}) - \xi_{i_0})\}\\
        & = P(t_{i_0})+2\xi_{i_0}+ \min\{2(1-\beta)(P(t_{i_0})-2\xi_{i_0}),2\beta(P(t_{i_0}) - 2\xi_{i_0})\}\\
        &\geq P(t_{i_0})+2\xi_{i_0} + \min\{0,2(P(t_{i_0})-2\xi_{i_0})\}\\
        &=\min\{P(t_{i_0}),2(P(t_{i_0})-\xi_{i_0})\}+\min\{P(t_{i_0}),2\xi_{i_0}\},
    \end{split}
\end{equation}
we have
\begin{equation}
    \begin{split}
    &\sum_{t\in V}\min\{P(t),F_{\beta}(P|k)(t)\}+\sum_{t\in V}\min\{P(t),F_{\beta}(P|k^r)(t)\}\\
    & = 1-P(t_0)+2\beta(1-P(t_0))+ \min\{P(t_{i_0}),2(1-\beta)\xi_{i_0}+2\beta(P(t_{i_0}) - \xi_{i_0})\}\\
    &+\min\{P(t_{i_0}),2(1-\beta)(P(t_{i_0})-\xi_{i_0})+2\beta\xi_{i_0}\}\\
    &\geq \sum_{t\in V}\min\{P(t),F_{PR}(P|k)(t)\}+\sum_{t\in V}\min\{P(t),F_{PR}(P|k^r)(t)\}+2\beta-2\beta P(t_{i_0})\\
    &\geq \sum_{t\in V}\min\{P(t),F_{PR}(P|k)(t)\}+\sum_{t\in V}\min\{P(t),F_{PR}(P|k^r)(t)\}+2\beta-2\beta \max_{t\in V}P(t).
    \end{split}
\end{equation}
Thus,
\begin{equation}
    \begin{split}
    \mathbb{D}(P,F_{\beta}) &= 1-\bbE_{k}[\sum_{t\in V}\min\{P(t),F_{\beta}(P|k)(t)\}]\\
    & = 1-\frac{1}{2}\bbE_{k}[\sum_{t\in V}\min\{P(t),F_{\beta}(P|k)(t)\}+\sum_{t\in V}\min\{P(t),F_{\beta}(P|k^r)(t)\}]\\
    & \leq 1-\frac{1}{2}\bbE_{k}[\sum_{t\in V}\min\{P(t),F_{PR}(P|k)(t)\}+\sum_{t\in V}\min\{P(t),F_{PR}(P|k^r)(t)\}+2\beta-2\beta \max_{t\in V}P(t)]\\
        &= \mathbb{D}(P,F_{PR}) -\beta(1-\max_{t\in V}P(t)).
    \end{split}
\end{equation}

\textbf{Part 2.} We then show $\forall P\in\cP$, if $\beta_1\leq\beta_2$, then $\mathbb{D}(P,F_{\beta_1})\geq \mathbb{D}(P,F_{\beta_2})$. 
Consider $\mathbb{D}(P,F_{\beta_1}) - \mathbb{D}(P,F_{\beta_2})$, we have
\begin{equation}\label{eqn:5.4part2}
    \begin{split}
        &\mathbb{D}(P,F_{\beta_1}) - \mathbb{D}(P,F_{\beta_2})\\
        =&\bbE_{k}[\sum_{t\in V}\min\{P(t),F_{\beta_2}(P|k)(t)\}]-\bbE_{k}[\sum_{t\in V}\min\{P(t),F_{\beta_1}(P|k)(t)\}]\\
        =&\frac{1}{2}\bbE_{k}\left[\sum_{t\in V}\min\{P(t),F_{\beta_2}(P|k)(t)\}+\sum_{t\in V}\min\{P(t),F_{\beta_2}(P|k^r)(t)\}\right.\\
        -&\left. \sum_{t\in V}\min\{P(t),F_{\beta_1}(P|k)(t)\}-\sum_{t\in V}\min\{P(t),F_{\beta_1}(P|k^r)(t)\}
        \right]
    \end{split}
\end{equation}

From the similar analysis as Part 1 we have for $F_{\beta_1}$,
$$\sum_{t\in V}\min\{P(t),F_{\beta_1}(P|k)(t)\} = \sum_{i=i_0+1}^NP(t_i)+2\beta_1\sum_{i=1}^{i_0-1}P(t_i)+\min\{P(t_{i_0}),2(1-\beta_1)\xi_{i_0}+2\beta_1(P(t_{i_0}) - \xi_{i_0})\},$$
and 

$$\sum_{t\in V}\min\{P(t),F_{\beta_1}(P|k^r)(t)\} = \sum_{i=1}^{i_0-1}P(t_i)+2\beta_1\sum_{i=i_0+1}^NP(t_i)+\min\{P(t_{i_0}),2(1-\beta_1)(P(t_{i_0})-\xi_{i_0})+2\beta_1\xi_{i_0}\}.$$

for $F_{\beta_2}$,
$$\sum_{t\in V}\min\{P(t),F_{\beta_2}(P|k)(t)\} = \sum_{i=i_0+1}^NP(t_i)+2\beta_2\sum_{i=1}^{i_0-1}P(t_i)+\min\{P(t_{i_0}),2(1-\beta_2)\xi_{i_0}+2\beta_2(P(t_{i_0}) - \xi_{i_0})\},$$
and
$$\sum_{t\in V}\min\{P(t),F_{\beta_2}(P|k^r)(t)\} = \sum_{i=1}^{i_0-1}P(t_i)+2\beta_2\sum_{i=i_0+1}^NP(t_i)+\min\{P(t_{i_0}),2(1-\beta_2)(P(t_{i_0})-\xi_{i_0})+2\beta_2\xi_{i_0}\}.$$

When $\beta_2\geq\beta_1$\begin{equation}
    \begin{split}
        &\min\{P(t_{i_0}),2(1-\beta_2)\xi_{i_0}+2\beta_2(P(t_{i_0}) - \xi_{i_0})\}+\min\{P(t_{i_0}),2(1-\beta_2)(P(t_{i_0})-\xi_{i_0})+2\beta_2\xi_{i_0}\}\\
        & = P(t_{i_0}) + \min\{2(1-\beta_2)(P(t_{i_0})-\xi_{i_0})+2\beta_2\xi_{i_0},2(1-\beta_2)\xi_{i_0}+2\beta_2(P(t_{i_0}) - \xi_{i_0})\}\\
        & = P(t_{i_0})+2\xi_{i_0}+ \min\{2(1-\beta_2)(P(t_{i_0})-2\xi_{i_0}),2\beta_2(P(t_{i_0}) - 2\xi_{i_0})\}\\
        &\geq P(t_{i_0})+2\xi_{i_0} + \min\{2(1-\beta_1)(P(t_{i_0})-2\xi_{i_0}),2\beta_1(P(t_{i_0}) - 2\xi_{i_0})\}\\
        &=\min\{P(t_{i_0}),2(1-\beta_1)\xi_{i_0}+2\beta_1(P(t_{i_0}) - \xi_{i_0})\}+\min\{P(t_{i_0}),2(1-\beta_1)(P(t_{i_0})-\xi_{i_0})+2\beta_1\xi_{i_0}\},
    \end{split}
\end{equation}
Thus, 
\begin{equation}
    \begin{split}
    &\sum_{t\in V}\min\{P(t),F_{\beta_2}(P|k)(t)\}+\sum_{t\in V}\min\{P(t),F_{\beta_2}(P|k^r)(t)\}\\
    & = 1-P(t_0)+2\beta_2(1-P(t_0))+ \min\{P(t_{i_0}),2(1-\beta_2)\xi_{i_0}+2\beta_2(P(t_{i_0}) - \xi_{i_0})\}\\
    &+\min\{P(t_{i_0}),2(1-\beta_2)(P(t_{i_0})-\xi_{i_0})+2\beta_2\xi_{i_0}\}\\
    &\geq \sum_{t\in V}\min\{P(t),F_{\beta_1}(P|k)(t)\}+\sum_{t\in V}\min\{P(t),F_{\beta_1}(P|k^r)(t)\}+2(\beta_2-\beta_1)(1-P(t_{i_0}))\\
    &\geq \sum_{t\in V}\min\{P(t),F_{\beta_1}(P|k)(t)\}+\sum_{t\in V}\min\{P(t),F_{\beta_1}(P|k^r)(t)\}+2(\beta_2-\beta_1)(1-\max_{t\in V}{P(t)}).
    \end{split}
\end{equation}
Combining with Equation~\ref{eqn:5.4part2} we have:
\begin{equation}
    \begin{split}
        &\mathbb{D}(P,F_{\beta_1}) - \mathbb{D}(P,F_{\beta_2})\\
        =&\bbE_{k}[\sum_{t\in V}\min\{P(t),F_{\beta_2}(P|k)(t)\}]-\bbE_{k}[\sum_{t\in V}\min\{P(t),F_{\beta_1}(P|k)(t)\}]\\
        =&\frac{1}{2}\bbE_{k}\left[\sum_{t\in V}\min\{P(t),F_{\beta_2}(P|k)(t)\}+\sum_{t\in V}\min\{P(t),F_{\beta_2}(P|k^r)(t)\}\right.\\
        -&\left. \sum_{t\in V}\min\{P(t),F_{\beta_1}(P|k)(t)\}-\sum_{t\in V}\min\{P(t),F_{\beta_1}(P|k^r)(t)\}
        \right]\\
        \geq& (\beta_2-\beta_1)(1-\max_{t\in V}{P(t)})\geq 0
    \end{split}
\end{equation}
Therefore, $\mathbb{D}(P,F_{\beta_1}) \geq \mathbb{D}(P,F_{\beta_2})$
\end{proof}

\subsection{Proof of Definition~\ref{def:model agnostic detection}}
\begin{proof}
    We prove the concentration bound in Definition~\ref{def:model agnostic detection}: $\Pr(S(\x_{1:n})-\mathbb{E}_{H_0}[S(\x_{1:n})]>t\sqrt{n}|H_0)\leq\exp(-2t^2)$.
Since the range of the sigmoid function is in $[0,1]$, by Hoeffding's inequality, for each random score $s(x_i)$, we have
\begin{equation}
    \Pr(\frac{1}{n}\sum_{i=1}^n s(x_i)-\mathbb{E}_{H_0}[\frac{1}{n}\sum_{i=1}^n s(x_i)]>t|H_0)\leq e^{-\frac{2t^2}{n}}
\end{equation}
Replace $t$ by $\frac{t}{\sqrt{n}}$ we have
\begin{equation}
    \Pr(\sum_{i=1}^n s(x_i)-\mathbb{E}_{H_0}[\sum_{i=1}^n s(x_i)]>t\sqrt{n}|H_0)\leq e^{-2t^2}
\end{equation}
    
\end{proof}

\section{Detailed Experiment Setup}\label{sec:detailed_experiment_setup}

\subsection{Experiment Setup}
We evaluate the distortion-free performance of various watermark models within two seq2seq applications: text summarization and text generation. The experiments leverage the Huggingface library \citep{wolf2019huggingface}, a popular framework for model development and sharing in the NLP community. All tests are conducted on 8 NVIDIA A6000 GPUs, each with 48GB of memory.

We focus on three seq2seq tasks in our experiments: machine translation, text summarization and text generation. For the machine translation task, we focus on English-to-Romanian translation. We employ the Multilingual BART (MBart) model \citep{liu2020multilingual} on the WMT’14 En-Ro corpus. For text summarization, we employ the BART-large model~\citep{liu2020multilingual} using the CNN-DM corpus dataset~\citep{hermann2015teaching}. For text generation, we follow the settings described by~\citep{kirchenbauer2023watermark}, using the LLaMA-2 model (7b, chat)~\citep{touvron2023llama2} with a random subset of the C4 dataset~\citep{raffel2020exploring}. All experiments are conducted with n-gram watermark key sampling ($n=5$). Additionally, we include the Soft watermark~\citep{kirchenbauer2023watermark} in our comparison, although it does not achieve step-wise distortion-free performance. Notably, when $\beta=0$, the Beta-watermark becomes identical to the permute-reweight watermark~\citep{hu2023unbiased}. 

\textbf{Machine Translation.} For the machine translation task, we utilize the WMT'14 English (En) to Romanian (Ro) dataset, comprising 1,999 examples in the test set. We employ the Multilingual Bart (MBart) model \citep{liu2020multilingual} along with its official tokenizer.

\textbf{Text Summarization.} For text summarization, we utilize the test set from the CNN-DM corpus \citep{hermann2015teaching}, which contains 11,490 examples. We employ the BART-large model, which has 400 million parameters, and the LLaMA-2 model with 7 billion parameters.

\textbf{Text Generation.} In text generation, we adhere to the experimental setup described in \cite{kirchenbauer2023watermark}. We use a random subset of the C4 dataset for generation prompts. Our model selection includes the LLaMA-2, which has 7 billion parameters.

\textbf{Watermark Setup.} Our experiments primarily compare the beta-watermark with three other distortion-free watermarks: inversa-sampling, Gumbel-reparametrization, and permute-reweight. Additionally, we include the Soft watermark \citep{kirchenbauer2023watermark} in our comparison. For beta-watermark, we explore various $\beta$ values from the set $\{0, 0.05, 0.1, 0.2, 0.3\}$. For the Soft watermark \citep{kirchenbauer2023watermark}, we investigate green list bias $\delta$ values from $\{0.5, 1.0, 1.5, 2.0\}$ with a fixed green list separator $\gamma=0.5$.
For n-gram key sampling, we consider the most recent 5 tokens as the texture key. For example, when generating $x_4$ in response to $(x_1, x_2, x_3)$, the texture key includes $(x_1, x_2, x_3)$, given only three tokens are available. Texture key history resets before generating each batch. For cipher generation, we use SHA-256 as the hash function and a 1024-bit random bitstrings as the secret key $\sk$, the watermark key is given by $k = (\sk,\x_{i-5,i-1})$. The permutation $\pi$ is sampled using $\text{hash}(k)$ as the random seed. We also compare beta-watermark with inverse-sampling watermark \cite{kuditipudi2023robust} and permute-reweight watermark \cite{hu2023unbiased,wu2023dipmark}, following the settings in their open-sourced code\footnote{\url{https://github.com/jthickstun/watermark}}\footnote{\url{https://github.com/xiaoniu-578fa6bff964d005/UnbiasedWatermark}}.

\textbf{Evaluation Metrics for Text Quality.} In this part, we detail the metrics used to evaluate text quality:
\begin{itemize}
    \item \textbf{ROUGE Score.} For the summarization task, we employ the ROUGE score \citep{lin2004rouge}, which measures the overlap of n-grams between the generated summaries and the reference texts to evaluate how effectively the summary captures the essential content.
    \item \textbf{BLEU score.} For the machine translation task, we rely on the BLEU score \citep{papineni2002bleu}, emphasizing the lexical similarity between machine-generated translations and human reference translations.
    \item \textbf{BERTScore.} BERTScore \cite{zhang2019bertscore} calculates the similarity between two sentences by summing the cosine similarities of their token embeddings. We utilize BERTScore-F1, BERTScore-Precision, and BERTScore-Recall for assessing both text summarization and machine translation tasks.
    \item \textbf{Perplexity.} Perplexity, a concept from information theory, measures how well a probability model or distribution predicts a sample. It is used to compare the performance of probability models, where a lower perplexity indicates a more predictive model. We apply perplexity to evaluate both text summarization and text generation tasks.
\end{itemize}

\textbf{Evaluation Metrics for Detecting Efficiency of Watermarks.} In this section, we present the metrics used to evaluate the detectability of watermarks:
\begin{itemize}
    \item \textbf{Type I and II Errors.} We employ the true positive rate (TPR), false positive rate (FPR), true negative rate (TNR), and false negative rate (FNR) to assess watermark detection across a mix of watermarked and non-watermarked sentences. The FPR measures the Type I error, which occurs when the null hypothesis is incorrectly rejected when it is actually true. The FNR measures the Type II error, where there is a failure to reject a false null hypothesis.
\end{itemize}
\section{Additional Experimental Results}\label{sec:add exp results}
In this section, we introduce the additional experiments conducted in our paper.

\textbf{Weakly Distortion-Free.}
The full results are presented in Table~\ref{tab:one-time}. This figure shows that compared to the model without watermarks, all weakly distortion-free watermarks exhibit no significant performance bias in text summarization and text generation tasks. However, for the Soft-watermark, a significant performance bias is observable as $\delta$ increases. Besides, we also include a comprehensive results for the combination of all PDA-rules and all three kinds of key sampling methods under text generation tasks. The results are presented in Table~\ref{tab:text generation full}. We also don't observe the distribution bias under the $\Delta$ metrics.

\textbf{Strongly Distortion-Free.}
The full results are displayed in Table~\ref{tab:multi-time full}, where we include all PDA-rule and key sampling method into comparison. From this table, it is evident that compared to the no watermark model, all weakly distortion-free watermarks demonstrate performance bias across all tasks. In contrast, the Beta-watermark exhibits less bias compared to other weakly distortion-free watermarks. Additionally, as $\beta$ increases, the distribution bias is further reduced, consistent with our theoretical analysis.

\begin{figure}
    \centering
    \includegraphics[width=0.5\linewidth]{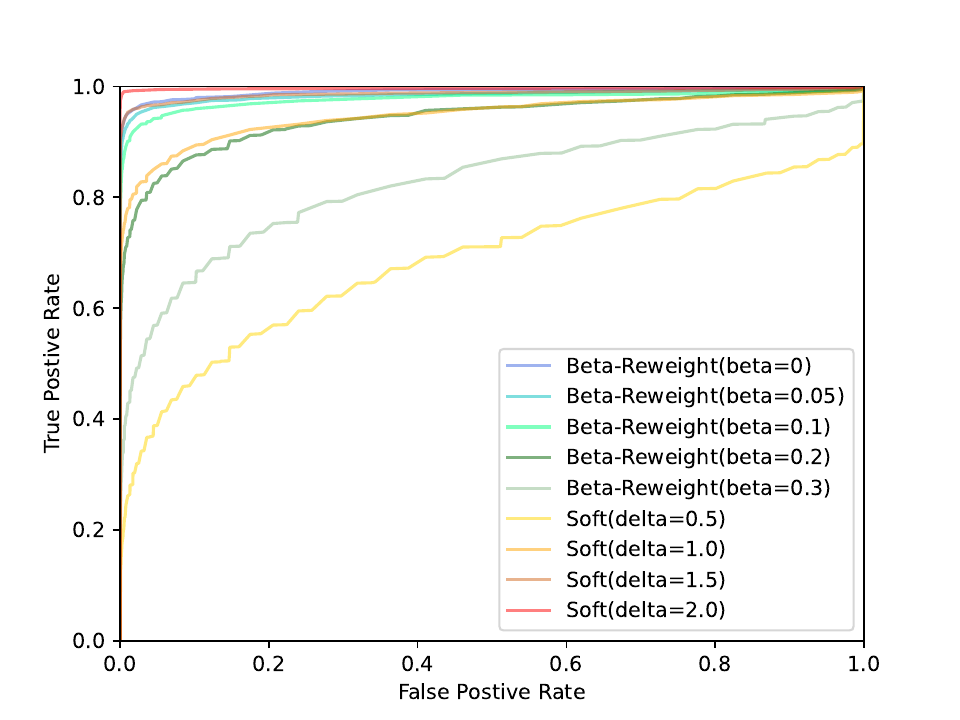}
    \caption{ROC curve of TPR vs FPR.}
    \label{fig:roc curve}
\end{figure}

\textbf{Detect efficiency.}
We compare the detection efficiency of beta-watermark with Soft-watermark on text generation tasks. 
In Figure~\ref{fig:roc curve}, we see that the ROC of beta watermark is comparable with the Soft-watermark~\citep{kirchenbauer2023watermark}. We also see that when $\beta$ increases, the detect efficiency decreases, this is because a larger $\beta$ introduces a smaller distribution bias into the watermarked distribution, thus reducing the watermark strength.

\begin{figure}\label{fig:trade-off}
    \centering
    \includegraphics[width=0.495\textwidth]{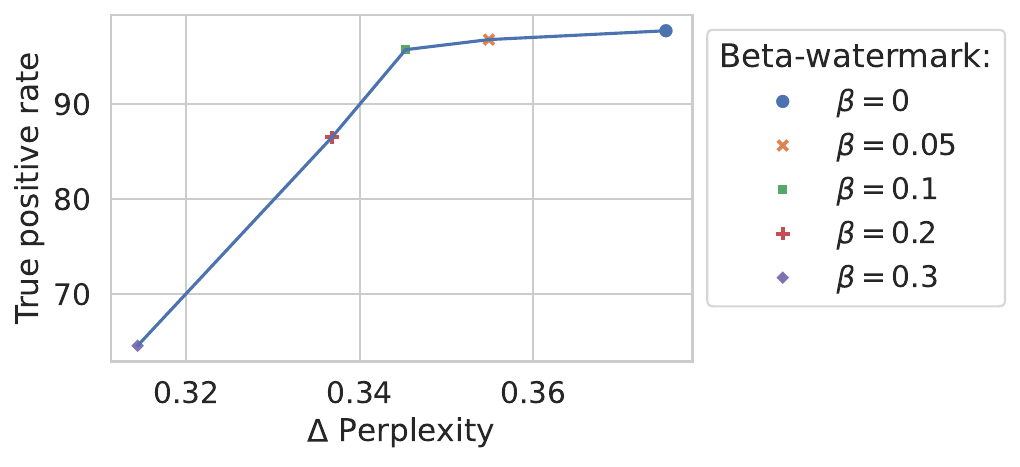}
    \includegraphics[width=0.495\textwidth]{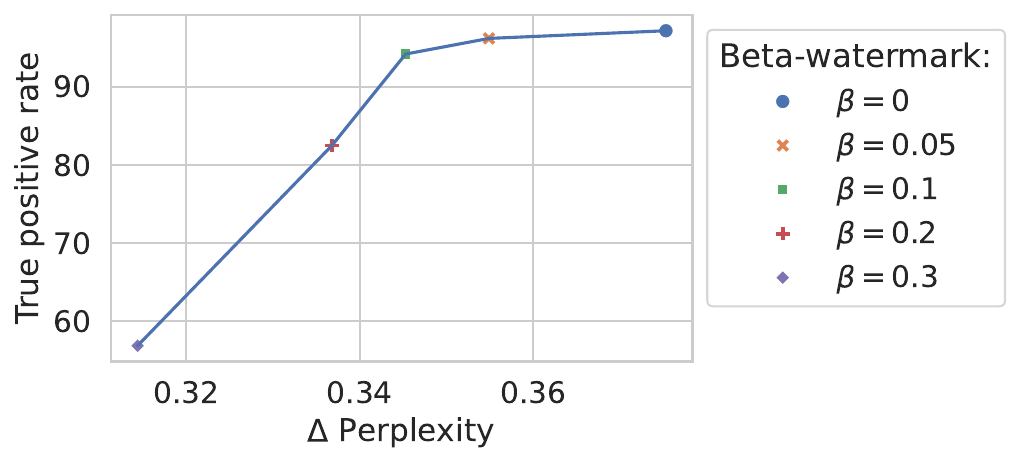}
    \includegraphics[width=0.495\textwidth]{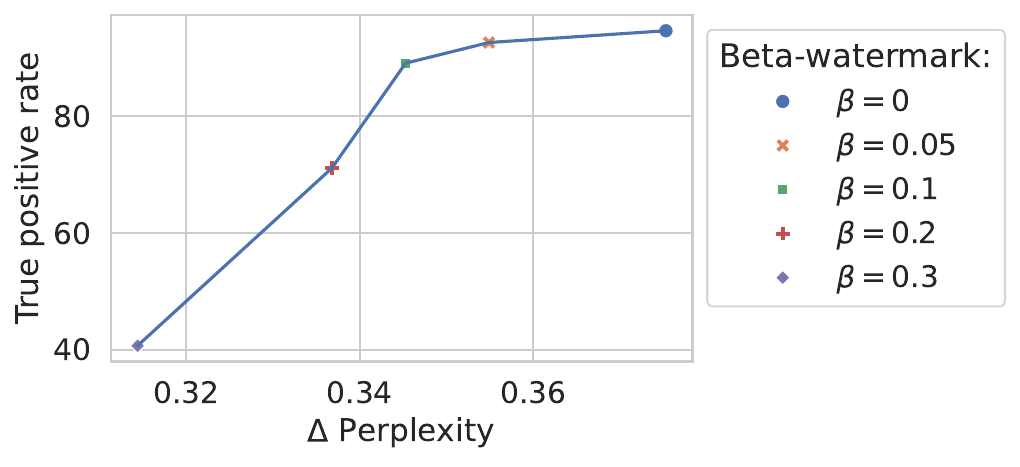}
    \includegraphics[width=0.495\textwidth]{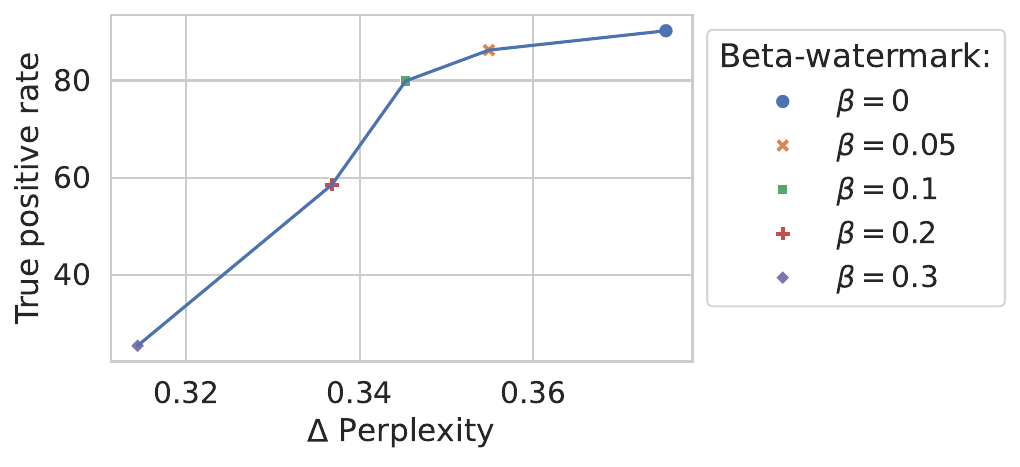}

    \caption{Trade-off between distribution bias and watermark strength under key collision. The TPR is measured under 10\% (\textbf{Top Left}), 5\% (\textbf{Top Right}), 1\% (\textbf{Bottom Left}), 0.1\% (\textbf{Bottom Right}) FPR. We can see $\Delta$ Perplexity (distribution bias) increase with the TPR. }
    \label{fig:trade-off full}
    \vspace{-0.2cm}
\end{figure}

We use the beta-watermark to illustrate the trade-off between watermark strength and distribution bias. As shown in Figure~\ref{fig:trade-off full}, with increasing values of $\beta$, the distribution bias decreases, but there is also a corresponding decrease in the true positive rate of watermark detection. This indicates that reducing the distribution bias of the watermark compromises its detectability.

\textbf{Robustness.} We assessed the robustness of the beta-watermark against random text paraphrase attacks \citep{kirchenbauer2023watermark}, where we modified 5\%, 10\%, 20\%, and 30\% of the tokens. The results, as detailed in Table~\ref{tab:robustness}, indicate that the beta-watermark maintains its robustness even with text modifications up to 30\%.

\begin{table}[]
\centering
\caption{Performance of different watermarks under one-time generation. For each prompt, only one response is generated.}
\label{tab:one-time}
\scalebox{0.85}{
\begin{tabular}{l|ccc|cc}
\hline
                                          & \multicolumn{3}{c|}{Text Summarization}                           & \multicolumn{2}{c}{Machine Translation} \\
                                          & BERT Score$\uparrow$ & ROUGE-1$\uparrow$ & Perplexity$\downarrow$ & BERT Score$\uparrow$  & BLEU $\uparrow$ \\ \hline
No Watermark                              & 0.3174±0.0885        & 0.3772±0.0962     & 6.4155±3.3009          & 0.2683±0.1967         & 10.8705±10.1914 \\ \hline
Beta-Reweight (\textbackslash{}beta=0)    & 0.3162±0.0871        & 0.3758±0.0961     & 6.3810±3.2753          & 0.2669±0.1966         & 10.6208±9.5880  \\
Beta-Reweight (\textbackslash{}beta=0.05) & 0.3171±0.0877        & 0.3760±0.0952     & 6.3986±3.2142          & 0.2683±0.1907         & 10.6511±10.1191 \\
Beta-Reweight (\textbackslash{}beta=0.1)  & 0.3169±0.0873        & 0.3762±0.0965     & 6.4250±3.2944          & 0.2687±0.1962         & 10.9058±10.5317 \\
Beta-Reweight (\textbackslash{}beta=0.2)  & 0.3184±0.0883        & 0.3771±0.0966     & 6.3889±3.2144          & 0.2641±0.1947         & 10.9852±10.7563 \\
Beta-Reweight (\textbackslash{}beta=0.3)  & 0.3167±0.0869        & 0.3764±0.0954     & 6.3972±3.2855          & 0.2668±0.1907         & 10.7865±9.8656  \\
Inverse-sampling                          & 0.3182±0.0876        & 0.3772±0.0964     & 6.3377±3.1274          & 0.2894±0.1869         & 11.6892±10.5368 \\
Gumbel-reparametrization                  & 0.3171±0.0868        & 0.3763±0.0961     & 6.3538±3.2221          & 0.3065±0.1875         & 11.8670±10.6599 \\ \hline
Soft($\delta$=0.5)                        & 0.3152±0.0862        & 0.3746±0.0949     & 6.4894±3.2453          & 0.2541±0.1950         & 10.3546±9.7336  \\
Soft($\delta$=1.0)                        & 0.3125±0.0856        & 0.3724±0.0937     & 6.8647±3.4364          & 0.2241±0.1922         & 9.5412±9.0065   \\
Soft($\delta$=1.5)                        & 0.3067±0.0825        & 0.3673±0.0917     & 7.4633±3.5928          & 0.1876±0.1891         & 8.5556±8.5925   \\
Soft($\delta$=2.0)                        & 0.2996±0.0805        & 0.3605±0.0899     & 8.4847±4.1598          & 0.1380±0.1750         & 6.9994±6.7528   \\ \hline
\end{tabular}
}
\end{table}
\begin{table}[]
\caption{Performance of different watermarks under one-time generation for text generation tasks. For each prompt, only one response is generated}
\label{tab:text generation full}
\scalebox{0.55}{
\begin{tabular}{ll|ccccccc}
\toprule
PDA-rule                       & Watermark key    & bertscore.precision & bertscore.recall & bertscore.f1  & ppl           & rouge1        & rouge2        & rougeL        \\ \midrule
                               & fixed key set    & 0.3062±0.0954       & 0.3279±0.1019    & 0.3170±0.0880 & 6.4090±3.2113 & 0.3764±0.0960 & 0.1324±0.0808 & 0.2377±0.0793 \\
$\beta$-reweight($\beta$=0)    & n-gram hashing   & 0.3048±0.0949       & 0.3276±0.1010    & 0.3162±0.0871 & 6.3810±3.2753 & 0.3758±0.0961 & 0.1314±0.0798 & 0.2372±0.0785 \\
                               & position hashing & 0.3050±0.0951       & 0.3271±0.1010    & 0.3160±0.0874 & 6.4285±3.2815 & 0.3759±0.0952 & 0.1315±0.0798 & 0.2374±0.0791 \\ \midrule
                               & fixed key set    & 0.3061±0.0953       & 0.3289±0.1026    & 0.3174±0.0884 & 6.3903±3.3533 & 0.3764±0.0964 & 0.1327±0.0806 & 0.2385±0.0801 \\
$\beta$-reweight($\beta$=0.05) & n-gram hashing   & 0.3058±0.0944       & 0.3286±0.1021    & 0.3171±0.0877 & 6.3986±3.2142 & 0.3760±0.0952 & 0.1320±0.0797 & 0.2375±0.0785 \\
                               & position hashing & 0.3058±0.0951       & 0.3283±0.1021    & 0.3170±0.0876 & 6.4043±3.3037 & 0.3763±0.0959 & 0.1326±0.0797 & 0.2385±0.0789 \\ \midrule
                               & fixed key set    & 0.3055±0.0948       & 0.3279±0.1014    & 0.3166±0.0873 & 6.4143±3.3500 & 0.3765±0.0956 & 0.1324±0.0795 & 0.2380±0.0785 \\
$\beta$-reweight($\beta$=0.1)  & n-gram hashing   & 0.3054±0.0950       & 0.3285±0.1015    & 0.3169±0.0873 & 6.4250±3.2944 & 0.3762±0.0965 & 0.1327±0.0801 & 0.2377±0.0785 \\
                               & position hashing & 0.3060±0.0954       & 0.3285±0.1008    & 0.3172±0.0875 & 6.4214±3.2642 & 0.3762±0.0952 & 0.1322±0.0785 & 0.2382±0.0780 \\ \midrule
                               & fixed key set    & 0.3068±0.0952       & 0.3296±0.1020    & 0.3181±0.0878 & 6.4131±3.3820 & 0.3778±0.0960 & 0.1337±0.0806 & 0.2395±0.0799 \\
$\beta$-reweight($\beta$=0.2)  & n-gram hashing   & 0.3068±0.0958       & 0.3302±0.1026    & 0.3184±0.0883 & 6.3889±3.2144 & 0.3771±0.0966 & 0.1334±0.0811 & 0.2392±0.0794 \\
                               & position hashing & 0.3057±0.0949       & 0.3283±0.1025    & 0.3169±0.0880 & 6.3685±3.2764 & 0.3765±0.0963 & 0.1323±0.0800 & 0.2383±0.0794 \\ \midrule
                               & fixed key set    & 0.3053±0.0955       & 0.3280±0.1018    & 0.3166±0.0878 & 6.3878±3.1945 & 0.3763±0.0954 & 0.1319±0.0799 & 0.2376±0.0788 \\
$\beta$-reweight($\beta$=0.3)  & n-gram hashing   & 0.3052±0.0949       & 0.3284±0.1006    & 0.3167±0.0869 & 6.3972±3.2855 & 0.3764±0.0954 & 0.1325±0.0799 & 0.2379±0.0784 \\
                               & position hashing & 0.3066±0.0952       & 0.3288±0.1018    & 0.3176±0.0876 & 6.3845±3.2077 & 0.3771±0.0963 & 0.1327±0.0798 & 0.2385±0.0787 \\ \midrule
                               & fixed key set    & 0.3011±0.0953       & 0.3277±0.1016    & 0.3143±0.0875 & 6.6430±3.5498 & 0.3746±0.0959 & 0.1309±0.0797 & 0.2361±0.0793 \\
Gumbel-reparametrization       & n-gram hashing   & 0.3060±0.0942       & 0.3284±0.1011    & 0.3171±0.0868 & 6.3538±3.2221 & 0.3763±0.0961 & 0.1321±0.0797 & 0.2376±0.0788 \\
                               & position hashing & 0.3047±0.0958       & 0.3267±0.1019    & 0.3156±0.0881 & 6.4877±3.4127 & 0.3755±0.0957 & 0.1317±0.0800 & 0.2380±0.0790 \\ \midrule
                               & fixed key set    & 0.3063±0.0942       & 0.3297±0.1014    & 0.3179±0.0870 & 6.1846±3.1150 & 0.3777±0.0960 & 0.1334±0.0802 & 0.2391±0.0793 \\
Inverse-sampling               & n-gram hashing   & 0.3064±0.0953       & 0.3302±0.1018    & 0.3182±0.0876 & 6.3377±3.1274 & 0.3772±0.0964 & 0.1328±0.0809 & 0.2390±0.0799 \\
                               & position hashing & 0.3075±0.0962       & 0.3326±0.1022    & 0.3199±0.0881 & 6.2007±3.0213 & 0.3796±0.0960 & 0.1344±0.0813 & 0.2404±0.0802 \\ \midrule
No Watermark                   & NA               & 0.3058±0.0959       & 0.3293±0.1026    & 0.3174±0.0885 & 6.4155±3.3009 & 0.3772±0.0962 & 0.1328±0.0806 & 0.2388±0.0799 \\ \midrule
Soft($\delta$=0.5)             & n-gram hashing   & 0.3013±0.0941       & 0.3294±0.1005    & 0.3152±0.0862 & 6.4894±3.2453 & 0.3746±0.0949 & 0.1310±0.0781 & 0.2362±0.0776 \\
Soft($\delta$=1.0)             & n-gram hashing   & 0.2956±0.0928       & 0.3296±0.0999    & 0.3125±0.0856 & 6.8647±3.4364 & 0.3724±0.0937 & 0.1279±0.0769 & 0.2328±0.0764 \\
Soft($\delta$=1.5)             & n-gram hashing   & 0.2858±0.0906       & 0.3280±0.0968    & 0.3067±0.0825 & 7.4633±3.5928 & 0.3673±0.0917 & 0.1229±0.0731 & 0.2271±0.0724 \\
Soft($\delta$=2.0)             & n-gram hashing   & 0.2751±0.0879       & 0.3246±0.0953    & 0.2996±0.0805 & 8.4847±4.1598 & 0.3605±0.0899 & 0.1158±0.0698 & 0.2207±0.0695 \\ \bottomrule
\end{tabular}
\vspace{5cm}
}
\\ 

\scalebox{0.55}{
\begin{tabular}{ll|ccccccc}
\toprule
PDA-rules                      & Watermark key    & $\Delta$ bertscore.precision & $\Delta$ bertscore.recall & $\Delta$ bertscore.f1 & $\Delta$ ppl  & $\Delta$ rouge1 & $\Delta$ rouge2 & $\Delta$ rougeL \\ \midrule
                               & fixed key set    & 0.0694±0.0564                & 0.0674±0.0577             & 0.0625±0.0520         & 2.7242±2.8964 & 0.0700±0.0549   & 0.0585±0.0517   & 0.0606±0.0519   \\
$\beta$-reweight($\beta$=0)    & n-gram hashing   & 0.0700±0.0561                & 0.0672±0.0567             & 0.0626±0.0513         & 2.7165±2.9231 & 0.0703±0.0560   & 0.0582±0.0517   & 0.0605±0.0519   \\
                               & position hashing & 0.0701±0.0565                & 0.0679±0.0575             & 0.0630±0.0518         & 2.7533±2.9858 & 0.0698±0.0554   & 0.0584±0.0521   & 0.0611±0.0533   \\ \midrule
                               & fixed key set    & 0.0701±0.0570                & 0.0678±0.0569             & 0.0630±0.0519         & 2.7436±3.0276 & 0.0709±0.0550   & 0.0588±0.0521   & 0.0617±0.0527   \\
$\beta$-reweight($\beta$=0.05) & n-gram hashing   & 0.0700±0.0567                & 0.0679±0.0573             & 0.0631±0.0519         & 2.7419±2.9226 & 0.0701±0.0554   & 0.0583±0.0517   & 0.0606±0.0522   \\
                               & position hashing & 0.0703±0.0566                & 0.0685±0.0577             & 0.0631±0.0521         & 2.7540±2.9807 & 0.0713±0.0560   & 0.0590±0.0524   & 0.0616±0.0522   \\ \midrule
                               & fixed key set    & 0.0695±0.0566                & 0.0674±0.0573             & 0.0623±0.0520         & 2.7563±3.0299 & 0.0693±0.0557   & 0.0580±0.0520   & 0.0608±0.0526   \\
$\beta$-reweight($\beta$=0.1)  & n-gram hashing   & 0.0696±0.0563                & 0.0676±0.0567             & 0.0626±0.0515         & 2.7640±2.9893 & 0.0701±0.0558   & 0.0579±0.0516   & 0.0605±0.0520   \\
                               & position hashing & 0.0703±0.0566                & 0.0676±0.0571             & 0.0630±0.0518         & 2.7559±2.9446 & 0.0698±0.0555   & 0.0583±0.0513   & 0.0610±0.0515   \\ \midrule
                               & fixed key set    & 0.0695±0.0560                & 0.0673±0.0570             & 0.0625±0.0512         & 2.7507±3.0184 & 0.0706±0.0553   & 0.0589±0.0524   & 0.0610±0.0525   \\
$\beta$-reweight($\beta$=0.2)  & n-gram hashing   & 0.0698±0.0566                & 0.0679±0.0571             & 0.0629±0.0517         & 2.7376±2.9355 & 0.0699±0.0558   & 0.0589±0.0525   & 0.0607±0.0518   \\
                               & position hashing & 0.0699±0.0563                & 0.0688±0.0587             & 0.0632±0.0526         & 2.7001±2.9368 & 0.0697±0.0563   & 0.0584±0.0529   & 0.0608±0.0532   \\ \midrule
                               & fixed key set    & 0.0706±0.0568                & 0.0680±0.0575             & 0.0631±0.0520         & 2.7242±2.9031 & 0.0701±0.0562   & 0.0581±0.0519   & 0.0608±0.0528   \\
$\beta$-reweight($\beta$=0.3)  & n-gram hashing   & 0.0705±0.0564                & 0.0679±0.0570             & 0.0633±0.0515         & 2.7466±2.9944 & 0.0701±0.0552   & 0.0585±0.0514   & 0.0609±0.0527   \\
                               & position hashing & 0.0696±0.0559                & 0.0673±0.0565             & 0.0622±0.0510         & 2.7271±2.9034 & 0.0693±0.0552   & 0.0576±0.0507   & 0.0602±0.0513   \\ \midrule
                               & fixed key set    & 0.0700±0.0572                & 0.0679±0.0578             & 0.0629±0.0524         & 2.8303±3.0803 & 0.0706±0.0561   & 0.0579±0.0523   & 0.0616±0.0530   \\
Gumbel-reparametrization       & n-gram hashing   & 0.0694±0.0561                & 0.0678±0.0574             & 0.0625±0.0517         & 2.7221±2.9595 & 0.0708±0.0555   & 0.0588±0.0520   & 0.0607±0.0524   \\
                               & position hashing & 0.0702±0.0573                & 0.0682±0.0585             & 0.0630±0.0530         & 2.7680±3.0449 & 0.0702±0.0563   & 0.0593±0.0529   & 0.0615±0.0539   \\ \midrule
                               & fixed key set    & 0.0692±0.0555                & 0.0661±0.0564             & 0.0618±0.0508         & 2.6649±2.8626 & 0.0695±0.0556   & 0.0580±0.0516   & 0.0608±0.0520   \\
Inverse-sampling               & n-gram hashing   & 0.0697±0.0565                & 0.0674±0.0567             & 0.0625±0.0516         & 2.7131±2.8903 & 0.0705±0.0557   & 0.0581±0.0521   & 0.0603±0.0523   \\
                               & position hashing & 0.0704±0.0559                & 0.0677±0.0579             & 0.0628±0.0517         & 2.6266±2.8591 & 0.0698±0.0559   & 0.0583±0.0519   & 0.0612±0.0526   \\ \midrule
Baseline                       & NA               & 0.0701±0.0560                & 0.0674±0.0570             & 0.0628±0.0513         & 2.7535±2.9630 & 0.0707±0.0558   & 0.0583±0.0522   & 0.0613±0.0527   \\ \midrule
Soft($\delta$=0.5)             & n-gram hashing   & 0.0700±0.0569                & 0.0677±0.0576             & 0.0627±0.0519         & 2.7403±2.9348 & 0.0700±0.0553   & 0.0581±0.0507   & 0.0606±0.0521   \\
Soft($\delta$=1.0)             & n-gram hashing   & 0.0692±0.0558                & 0.0666±0.0562             & 0.0616±0.0505         & 2.8607±3.0746 & 0.0688±0.0543   & 0.0569±0.0501   & 0.0595±0.0511   \\
Soft($\delta$=1.5)             & n-gram hashing   & 0.0704±0.0564                & 0.0661±0.0557             & 0.0613±0.0508         & 3.0427±3.1473 & 0.0688±0.0550   & 0.0566±0.0505   & 0.0593±0.0516   \\
Soft($\delta$=2.0)             & n-gram hashing   & 0.0736±0.0587                & 0.0669±0.0560             & 0.0635±0.0517         & 3.6349±3.6255 & 0.0699±0.0552   & 0.0576±0.0509   & 0.0601±0.0517   \\ \bottomrule
\end{tabular}
}
\end{table}

\begin{table}[]
\centering
\caption{Performance of different watermarks under multi-time generations. We randomly selected 1000 prompts and generated 100 responses for each. We use F1 scores of BERTScore and scale BERTScore and ROUGE-1 with a factor of 100.}
\label{tab:multi-time full}
\scalebox{0.55}{
\begin{tabular}{ll|lllllll}
\toprule
PDA-rule                       & Watermark key    & $\Delta$ bertscore.precision & $\Delta$ bertscore.recall & $\Delta$ bertscore.f1 & $\Delta$ ppl  & $\Delta$ rouge1 & $\Delta$ rouge2 & $\Delta$ rougeL \\ \midrule
                               & fixed key set    & 0.0070±0.0056                & 0.0066±0.0056             & 0.0062±0.0051         & 0.3123±0.2698 & 0.0071±0.0056   & 0.0062±0.0052   & 0.0062±0.0052   \\
$\beta$-reweight($\beta$=0)    & n-gram hashing   & 0.0095±0.0082                & 0.0097±0.0084             & 0.0090±0.0077         & 0.3753±0.3448 & 0.0093±0.0078   & 0.0091±0.0087   & 0.0100±0.0093   \\
                               & position hashing & 0.0092±0.0077                & 0.0095±0.0085             & 0.0086±0.0077         & 0.3711±0.3339 & 0.0091±0.0075   & 0.0088±0.0086   & 0.0099±0.0093   \\ \midrule
                               & fixed key set    & 0.0074±0.0060                & 0.0070±0.0060             & 0.0066±0.0055         & 0.3084±0.2880 & 0.0073±0.0060   & 0.0061±0.0055   & 0.0063±0.0056   \\
$\beta$-reweight($\beta$=0.05) & n-gram hashing   & 0.0091±0.0074                & 0.0092±0.0076             & 0.0084±0.0071         & 0.3549±0.3200 & 0.0085±0.0070   & 0.0084±0.0079   & 0.0092±0.0082   \\
                               & position hashing & 0.0087±0.0070                & 0.0089±0.0078             & 0.0083±0.0068         & 0.3488±0.3192 & 0.0084±0.0067   & 0.0081±0.0073   & 0.0089±0.0083   \\ \midrule
                               & fixed key set    & 0.0066±0.0052                & 0.0066±0.0054             & 0.0060±0.0047         & 0.3061±0.2696 & 0.0069±0.0055   & 0.0059±0.0052   & 0.0061±0.0051   \\
$\beta$-reweight($\beta$=0.1)  & n-gram hashing   & 0.0084±0.0071                & 0.0086±0.0070             & 0.0079±0.0065         & 0.3453±0.3214 & 0.0081±0.0068   & 0.0079±0.0073   & 0.0086±0.0078   \\
                               & position hashing & 0.0085±0.0069                & 0.0088±0.0073             & 0.0082±0.0066         & 0.3393±0.3195 & 0.0085±0.0066   & 0.0077±0.0069   & 0.0084±0.0074   \\ \midrule
                               & fixed key set    & 0.0072±0.0057                & 0.0069±0.0059             & 0.0065±0.0053         & 0.2960±0.2724 & 0.0073±0.0060   & 0.0062±0.0054   & 0.0062±0.0054   \\
$\beta$-reweight($\beta$=0.2)  & n-gram hashing   & 0.0076±0.0060                & 0.0078±0.0063             & 0.0070±0.0057         & 0.3368±0.3231 & 0.0077±0.0061   & 0.0071±0.0064   & 0.0076±0.0066   \\
                               & position hashing & 0.0078±0.0064                & 0.0077±0.0063             & 0.0072±0.0059         & 0.3229±0.2906 & 0.0076±0.0062   & 0.0070±0.0065   & 0.0077±0.0067   \\ \midrule
                               & fixed key set    & 0.0066±0.0054                & 0.0066±0.0055             & 0.0060±0.0048         & 0.3078±0.2786 & 0.0069±0.0056   & 0.0059±0.0052   & 0.0060±0.0051   \\
$\beta$-reweight($\beta$=0.3)  & n-gram hashing   & 0.0071±0.0056                & 0.0073±0.0058             & 0.0066±0.0052         & 0.3144±0.3015 & 0.0073±0.0060   & 0.0066±0.0056   & 0.0069±0.0058   \\
                               & position hashing & 0.0073±0.0059                & 0.0070±0.0060             & 0.0066±0.0054         & 0.3057±0.2991 & 0.0072±0.0059   & 0.0066±0.0057   & 0.0067±0.0058   \\ \midrule
                               & fixed key set    & 0.0080±0.0063                & 0.0074±0.0059             & 0.0070±0.0057         & 0.3744±0.3205 & 0.0079±0.0064   & 0.0067±0.0060   & 0.0069±0.0057   \\
Gumbel-reparametrization       & n-gram hashing   & 0.0480±0.0402                & 0.0461±0.0396             & 0.0428±0.0360         & 1.8892±1.8931 & 0.0488±0.0400   & 0.0409±0.0352   & 0.0427±0.0362   \\
                               & position hashing & 0.0494±0.0399                & 0.0485±0.0403             & 0.0442±0.0373         & 1.9935±2.4110 & 0.0512±0.0413   & 0.0423±0.0374   & 0.0442±0.0388   \\ \midrule
                               & fixed key set    & 0.0069±0.0056                & 0.0071±0.0061             & 0.0064±0.0054         & 0.3320±0.3066 & 0.0075±0.0061   & 0.0062±0.0057   & 0.0065±0.0052   \\
Inverse-sampling               & n-gram hashing   & 0.0486±0.0388                & 0.0481±0.0402             & 0.0439±0.0367         & 1.9380±2.0342 & 0.0499±0.0384   & 0.0403±0.0346   & 0.0428±0.0363   \\
                               & position hashing & 0.0503±0.0426                & 0.0469±0.0424             & 0.0448±0.0380         & 1.9095±2.2396 & 0.0491±0.0398   & 0.0422±0.0391   & 0.0441±0.0396   \\ \midrule
Baseline                       & NA               & 0.0068±0.0058                & 0.0067±0.0054             & 0.0062±0.0052         & 0.3028±0.2668 & 0.0070±0.0056   & 0.0060±0.0053   & 0.0060±0.0053   \\ \midrule
Soft($\delta$=0.5)             & n-gram hashing   & 0.0078±0.0063                & 0.0069±0.0057             & 0.0064±0.0053         & 0.3331±0.2965 & 0.0076±0.0061   & 0.0065±0.0056   & 0.0065±0.0056   \\
Soft($\delta$=1.0)             & n-gram hashing   & 0.0127±0.0096                & 0.0086±0.0074             & 0.0091±0.0077         & 0.5473±0.4023 & 0.0099±0.0083   & 0.0090±0.0080   & 0.0090±0.0078   \\
Soft($\delta$=1.5)             & n-gram hashing   & 0.0200±0.0129                & 0.0106±0.0093             & 0.0128±0.0104         & 1.1237±0.5868 & 0.0136±0.0110   & 0.0123±0.0110   & 0.0127±0.0107   \\
Soft($\delta$=2.0)             & n-gram hashing   & 0.0312±0.0175                & 0.0133±0.0125             & 0.0195±0.0146         & 2.0817±0.8216 & 0.0194±0.0149   & 0.0182±0.0156   & 0.0188±0.0149   \\ \bottomrule
\end{tabular}
}
\end{table}


\begin{table}
\centering
\caption{AUC score of different watermarks under varying attack strength $\epsilon$ on text generation task.}
\label{tab:robustness}
\scalebox{0.8}{
\begin{tabular}{l|ccccc}
\toprule
Beta-Reweight & $\epsilon$=0 & $\epsilon$=0.05 & $\epsilon$=0.1 & $\epsilon$=0.2 & $\epsilon$=0.3 \\ \midrule
$\beta$=0 & 0.9948 & 0.9901 & 0.9742 & 0.8848 & 0.7447 \\
$\beta$=0.05 & 0.9912 & 0.9846 & 0.9672 & 0.8724 & 0.7312 \\
$\beta$=0.1 & 0.9889 & 0.9785 & 0.9550 & 0.8558 & 0.7078 \\
$\beta$=0.2 & 0.9796 & 0.9598 & 0.9201 & 0.7983 & 0.6735 \\
$\beta$=0.3 & 0.9447 & 0.9047 & 0.8509 & 0.7289 & 0.6191 \\ \bottomrule
\end{tabular}
}
\end{table}

\clearpage
\section{Broader Impacts}\label{sec:broader impacts}

Machine learning models have profound impacts across various domains, demonstrating significant potential in both enhancing efficiencies and addressing complex challenges~\citep{yang2020prioritizing,yang2019lasso,wen2023feature,chakraborty2022using,cai2022asset,chen2024your,xu2017low,feng2018indexing} However, alongside these positive impacts, there are concerns about the integrity and security of machine learning applications~\citep{wu2023adversarial,wu2022retrievalguard,wu2023law,hong2024improving,pmlr-v202-hu23g,wang2023defending,wang2023distributionally}. Watermarking emerges as a pivotal technique in this context. It ensures the authenticity and ownership of digital media, and also can help people to distinguish AI generated contents.

\textbf{Positive Societal Impacts:}
\begin{itemize}

    \item Intellectual Property Protection: By developing robust watermarking methods, creators and developers can ensure their models are protected against unauthorized use or replication. This is especially vital as AI and machine learning models become more integral to industries like journalism, education, and entertainment, helping maintain economic incentives for innovation.

    \item Trust and Security: Enhanced watermarking techniques can help verify the authenticity and source of digital content, which is increasingly important in an era of deepfakes and misinformation. This can lead to greater trust in digital media and communications, as users can be more confident about the origin and integrity of the content they consume.

    \item Regulatory Compliance: Effective watermarking can assist in enforcing compliance with regulatory frameworks concerning data usage and copyrights. This could be particularly relevant in areas involving sensitive information, ensuring that content generation adheres to ethical standards and legal requirements.

\end{itemize}
\textbf{Negative Societal Impacts:}
\begin{itemize}
\item Privacy Concerns: While watermarking can secure content against misuse, it might also be used to track user behavior and preferences without their consent. If misused, such technologies could infringe on privacy and lead to surveillance issues, where every piece of generated content is potentially traceable back to its source.

\item Misuse in Propagating Bias: If watermarking techniques are used to proliferate content generated from biased models, it could further entrench and legitimize these biases. For example, if a biased language model is watermarked and widely distributed, it could contribute to the spread of discriminatory or prejudicial narratives under the guise of authenticated content.
\end{itemize}
\newpage
\end{document}